\documentclass[a4,12pt,eclepsf]{article}
\usepackage[dvips]{graphicx}
\usepackage{wrapfig, fancybox, fancyhdr, amsmath, amssymb, makeidx, 
comment, multirow, bm, here}
\usepackage[bf]{caption}

\newcommand{\BM}{\begin{minipage}}
\newcommand{\EM}{\end{minipage}}

\begin{document}

\renewcommand{\theequation}
{\thesection.\arabic{equation}}
\thispagestyle{empty}
\vspace*{15mm} 
\begin{center}
{\LARGE {\bf Baryon number current in}} \\[5mm]
{\LARGE {\bf holographic noncommutative QCD}} \\

\vspace*{15mm}

\renewcommand{\thefootnote}{\fnsymbol{footnote}}

{\Large Tadahito NAKAJIMA \footnotemark[1]\,\footnotemark[2]\,,
Yukiko OHTAKE \footnotemark[3] \\[3mm]
and \,Kenji SUZUKI \footnotemark[4]} \\
\vspace*{15mm}
$\footnotemark[1]$ {\it Department of Physics, Swansea University, 
Swansea SA2 8PP, United Kingdom} \\[2mm]
$\footnotemark[2]$ {\it College of Engineering, Nihon University, 
Fukushima 963-8642, Japan} \\[2mm]
$\footnotemark[3]$
{\it Toyama National College of Technology, Toyama 939-8630, Japan} \\[2mm]
$\footnotemark[4]$
{\it Department of Physics, Ochanomizu University, 
Tokyo 112-8610, Japan} \\[10mm]


{\bf Abstract} \\[10mm]

\end{center}

We consider the noncommutative deformation of the finite temperature holographic QCD (Sakai--Sugimoto) model in external electric and magnetic field and evaluate the effect of the noncommutaivity on the properties of the conductor-insulator phase transition associated with a baryon number current. Although the noncommutative deformation of the gauge theory does not change the phase structure with respect to the baryon number current, the transition temperature $T_{c}$, the transition electric field $e_{c}$ and magnetic field $b_{c}$ in the conductor-insurator phase transition depend on the noncommutativity parameter $\theta$. Namely, the noncommutativity of space coordinates has an influence on the shape of the phase diagram for the conductor-insurator phase transition. On the other hand, the allowed range of the noncommutativity parameter can be restricted by the reality condition of the constants of motion.

\clearpage

\setcounter{section}{0}
\section{Introduction}
\setcounter{page}{1}
\setcounter{equation}{0}

Noncommutative gauge theories (gauge theories on noncommutative Moyal space) can be realized as low energy theories of D-branes with Neveu-Schwarz-Neveu-Schwarz (NS-NS) $B$(two-form) field \cite{CDS, DH, AASJ, SJ, SW, DN, RJS}. The noncommutativity of space coordinates brings nontrivial properties on the gauge field theory at the quantum level. A remarkable phenomenon is so-called UV/IR mixing \cite{MRS}, where the ultraviolet (UV) and infrared (IR) degrees of freedom of the theory are mixed in a complicated non-trivial way. Although the noncommutative gauge theories have been studied extensively, it is hard to investigate them in the perturbative approach. Little is currently known of the non-perturbative properties of noncommutative gauge theories.

The noncommutative Yang--Mills theories have gravity duals whose near horizon region describes the noncommutative Yang--Mills theories in the limit of large $N_{c}$ and large coupling \cite{HI, MR, AOSJ}. Based on the generalized gauge/gravity (or AdS/CFT) duality, we can explore the non-perturbative aspects of the noncommutative gauge theories. For instance, the noncommutativity of space coordinates modifies the Wilson loop behavior \cite{dhar_kita, lee_sin, taka_naka-suzu} and glueball mass spectra \cite{NST}. The gravity duals of noncommutative gauge theories with matter in the fundamental representation have also been constructed by adding probe flavor branes \cite{APR}. Employing the gravity dual description of noncommutative gauge theories with flavor degrees of freedom we have been able to find the noncommutativity is also reflected in the flavor dynamics. For instance, the mass spectrum of mesons can be modified by the noncommutativity of space coordinates \cite{APR}.

Fundamental properties of quantum chromodynamics (QCD) at low energies are 
confinement and chiral symmetry breaking. The Sakai--Sugimoto model (a holographic QCD model with $\text{D4-D8-}\overline{\text{D8}}$-brane system) has been known to capture these properties of QCD at low energies \cite{SS1, SS2}. The holographic QCD models can be modified to introduce finite temperature. The phase of chiral symmetry breaking and restoration can be interpreted as configurations of probe branes in this model \cite{ASY, PS, PSZ}. The effect of the noncommutativity on the chiral phase transition have been examined by the noncommutative deformation of the holographic QCD model at finite temperature. The phase diagrams for the the chiral phase transition can be deformed by the noncommutativity of space coordinates \cite{TN_YO_KS}.

It has been shown that the large $N_{c}$ QCD at finite temperature has conductor and insulator phase associated with a baryon number current within a framework of the finite temperature Sakai--Sugimoto model in external electric and magnetic field \cite{BLL, CVJAK}, a la Karch--O'Bannon \cite{KOB}. This conductor-insulator phase transition is closely related to chiral phase transition in the finite temperature Sakai--Sugimoto model. This fact suggests the possibility that the phase diagrams for the conductor-insulator phase transition associated with a baryon number current can also be deformed by the noncommutativity of space coordinates.

We construct the noncommutative deformation of the finite temperature holographic QCD (Sakai--Sugimoto) model in external electric and magnetic field and evaluate the effect of the noncommutativity on the properties of the conductor-insulator phase transition associated with a baryon number current. As will be seen later, the baryon number current, the conductivity and the phase diagrams for the conductor-insulator phase transition can be deformed by the noncommutativity of space coordinates.\footnotemark[1] The Wess--Zumino term in the effective action of the probe branes plays the role of the noncommutative deformation on the properties of the conductor-insulator phase transition. 

\footnotetext[1]{
The noncommutative deformation on the conductivity associated with a baryon number current has been examined by \cite{MAA}. The response of the properties of the conductor-insulator phase transition associated with a baryon number current to NS-NS field has been examined by \cite{YSSSWX}}.

This paper is organized as follows. In section 2, we introduce the holographic QCD (Sakai--Sugimoto) model at finite temperature and discuss the features of the phase transition. Then we construct the noncommutative deformation of this model. In section 3, we investigate the response of the baryon number current to the external electric field and evaluate the noncommutative deformation of the baryon number current, the conductivity and the phase diagrams for the conductor-insulator phase transition. In section 4, we investigate the response to the external magnetic field and evaluate the noncommutative deformation of the phase diagrams. Section 5 is devoted to conclusions and discussions.

%
%

\section{Noncommutative deformation of the holographic QCD model at finite 
temperature} 

\setcounter{equation}{0}
\addtocounter{enumi}{1}

In this section, we consider a noncommutative deformation of the holographic 
QCD (Sakai--Sugimoto) model at finite temperature based on the prescription 
of \cite{APR}. The holographic QCD model is a gravity dual for a $4+1$ 
dimensional QCD with ${\rm U}({\rm N}_{f})_{L} \times {\rm U}({\rm N}_{f})_{R}$ global chiral symmetry whose symmetry is spontaneously broken \cite{SS1, SS2}. 
This model is a $\text{D4-D8-}\overline{\text{D8}}$-brane system consisting 
$S^{1}$ compactified $N_{c}$ D4-branes and $N_{f}$ 
$\text{D8-}\overline{\text{D8}}$-branes pairs transverse to the $S^{1}$. The 
near-horizon limit of the set of $N_{c}$ D4-branes solution compactified on 
$S^{1}$ takes the following form:
\begin{align}
ds^{2} &=\left(\frac{u}{R_{\rm D4}}\right)^{3/2}\Bigl(-(dt)^{2}+(dx^{1})^{2}
+ (dx^{2})^{2}+(dx^{3})^{2} + f_{K}(u)\,d\tau^{2} \Bigr) \nonumber \\
& + \left(\frac{R_{\rm D4}}{u}\right)^{3/2} 
\left(\frac{du^{2}}{f_{K}(u)}+u^{2}d\Omega_{4}^{2} \right)\,,\nonumber \\
& R_{\rm D4}^{3}=\pi g_{s}N_{c}l^{3}_{s}\,, \qquad 
f_{K}(u)=1-\frac{u_{K}{}^{3}}{u^{3}}\,,
\label{201}
\end{align}
where $u_{K}$ is a parameter, $u$ is the radial direction bounded from below by $u \geq u_{K}$, $\tau$ is compactified direction of the D4-brane world volume which is transverse to the $\text{D8-}\overline{\text{D8}}$-branes, $g_{s}$ and $l_{s}$ are the string coupling and the string length, respectively. The dilaton $\phi$ and the field strength $F_{4}$ of the RR 3-form $C_{3}$ are given by 
\begin{align}
e^{\phi}=g_{s}\left(\dfrac{u}{R_{\rm D4}}\right)^{3/4}\,, \qquad 
F_{4}=dC_{3}=\dfrac{2\pi N_{c}}{V_{4}}\epsilon_{4}\,,
\label{202}
\end{align}
where $V_{4}=8\pi^{2}/3$ is the volume of unit $S^{4}$ and $\epsilon_{4}$ is the corresponding volume form. In order to avoid a conical singularity at $u=u_{K}$, the $\tau$ direction should have a period of 
%
\begin{align}
\delta \tau 
= \dfrac{4\pi}{3}\left(\dfrac{R_{\rm D4}^{3}}{u_{K}}\right)^{1/2}
= 2\pi R = \dfrac{2\pi}{M_{\rm KK}}\,,
\label{203}
\end{align}
where $R$ is radius of $S^{1}$ and $M_{\rm KK}$ is the Kaluza--Klein mass. The parameter $u_{\rm K}$ is related to the Kaluza--Klein mass $M_{\rm KK}$ via the relation (\ref{203}). The five dimensional gauge coupling is expressed in terms of $g_{s}$ and $l_{s}$ as $g_{\rm YM}^{2}=(2\pi)^{2}g_{s}l_{s}$. The gravity description is valid for strong coupling $\lambda \gg R$, where as usual $\lambda = g_{\rm YM}^{2}N_{c}$ denotes the 't Hooft coupling. 

Next, we consider the probe D8-branes and anti D8-branes($\overline{\text{D8}}$-branes) which span the coordinates \,$t,\,x^{i}(i=1,2,3),\,\Omega_{4}$. They are treated as probes in the D4-brane background. The flavour degrees of freedom are introduced by strings stretching between the D4-branes and D8($\overline{\text{D8}}$)-branes. The D8-branes and $\overline{\text{D8}}$-branes are connected at $u=u_{0}$ as shown in Fig. 1. The connected configuration of the $\text{D8-}\overline{{\text{D8}}}$-branes indicates that the ${\rm U}({\rm N}_{f})_{L} \times {\rm U}({\rm N}_{f})_{R}$ global chiral symmetry is broken to a diagonal subgroup ${\rm U}({\rm N}_{f})$. We refer to the  connected configuration in the low temperature as the low-temperature phase. 

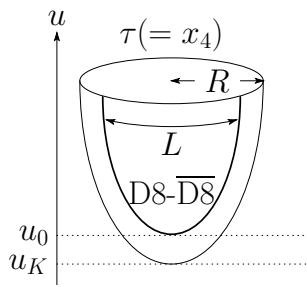
\begin{figure}[H]
\centering
\scalebox{0.6}{
{\unitlength 0.1in%
\begin{picture}(28.5500,23.9500)(6.4500,-30.3000)%
\put(13.0000,-7.0000){\makebox(0,0){\LARGE $u$}}%
%
\special{pn 8}%
\special{ar 2300 1250 800 200 0.0000000 6.2831853}%
%
\special{pn 8}%
\special{ar 2300 1250 800 1600 6.2831853 3.1415927}%
%
\special{pn 15}%
\special{ar 2300 1390 600 1200 6.2831853 3.1415927}%
\put(10.5000,-28.5000){\makebox(0,0){\LARGE $u_{K}$}}%
\put(11.0000,-26.0000){\makebox(0,0){\LARGE $u_{0}$}}%
\put(23.0000,-8.5000){\makebox(0,0){\LARGE $\tau(=x_{4})$}}%
\put(23.0000,-22.0000){\makebox(0,0){\LARGE ${\rm D8}\text{-}\overline{\rm D8}$}}%
%
\special{pn 8}%
\special{pa 1300 3030}%
\special{pa 1300 830}%
\special{fp}%
\special{sh 1}%
\special{pa 1300 830}%
\special{pa 1280 897}%
\special{pa 1300 883}%
\special{pa 1320 897}%
\special{pa 1300 830}%
\special{fp}%
%
\special{pn 8}%
\special{pa 1300 2600}%
\special{pa 3500 2600}%
\special{dt 0.045}%
%
\special{pn 8}%
\special{ar 2300 1450 800 200 0.7853982 2.3561945}%
%
\special{pn 8}%
\special{pa 1752 1596}%
\special{pa 1734 1591}%
\special{fp}%
\special{sh 1}%
\special{pa 1734 1591}%
\special{pa 1793 1628}%
\special{pa 1785 1605}%
\special{pa 1804 1590}%
\special{pa 1734 1591}%
\special{fp}%
%
\special{pn 8}%
\special{pa 2848 1596}%
\special{pa 2866 1591}%
\special{fp}%
\special{sh 1}%
\special{pa 2866 1591}%
\special{pa 2796 1590}%
\special{pa 2815 1605}%
\special{pa 2807 1628}%
\special{pa 2866 1591}%
\special{fp}%
\put(23.0000,-18.0000){\makebox(0,0){\LARGE $L$}}%
%
\special{pn 8}%
\special{pa 2900 1250}%
\special{pa 3100 1250}%
\special{fp}%
\special{sh 1}%
\special{pa 3100 1250}%
\special{pa 3033 1230}%
\special{pa 3047 1250}%
\special{pa 3033 1270}%
\special{pa 3100 1250}%
\special{fp}%
\special{pa 2500 1250}%
\special{pa 2300 1250}%
\special{fp}%
\special{sh 1}%
\special{pa 2300 1250}%
\special{pa 2367 1270}%
\special{pa 2353 1250}%
\special{pa 2367 1230}%
\special{pa 2300 1250}%
\special{fp}%
\put(27.0000,-12.5000){\makebox(0,0){\LARGE $R$}}%
%
\special{pn 8}%
\special{pa 1300 2850}%
\special{pa 3500 2850}%
\special{dt 0.045}%
\end{picture}}%
} \\
\caption{The $\text{D8-}\overline{\text{D8}}$-branes configurations at low temperature.} 
\label{LT}
\end{figure}

The holographic QCD model at finite temperature has been proposed in \cite{ASY, PS, PSZ}. In order to introduce a finite temperature $T$ in the model, we consider the Euclidean gravitational solution which is asymptotically equals to (\ref{201}) but with the compactification of Euclidean time direction $t_{E}$. In this solution the periodicity of $t_{E}$ is arbitrary and equals to $\beta=1/T$. Another solution with the same asymptotic is given by interchanging the role of $t_{E}$ and $\tau$ directions, 
\begin{align}
ds^{2} &=\left(\frac{u}{R_{\rm D4}}\right)^{3/2}
\Bigl(f_{T}(u)\,(dt_{E})^{2}+(dx^{1})^{2}
+ (dx^{2})^{2}+(dx^{3})^{2} + d\tau^{2} \Bigr) \nonumber \\
& + \left(\frac{R_{\rm D4}}{u}\right)^{3/2}
\left(\frac{du^{2}}{f_{T}(u)}+u^{2}d\Omega_{4}^{2} \right)\,,\nonumber \\
& R_{\rm D4}^{3}=\pi g_{s}N_{c}l^{3}_{s}\,, \qquad 
f_{T}(u)=1-\frac{u^{3}_{T}}{u^{3}}\,,
\label{204}
\end{align}
where $u_{T}$ is a parameter. The period of the compactified time direction is set to 
\begin{align}
\delta t_{E} = \dfrac{4\pi}{3}\left(\dfrac{R_{\rm D4}^{3}}{u_{T}}\right)^{1/2}
=\dfrac{1}{T}\,
\label{205}
\end{align}
to avoid a singularity at $u=u_{T}$. The parameter $u_{T}$ is related to the temperature $T$. The metric (\ref{201}) with the compactification of Euclidean time $t_{E}$ is dominant in the low temperature $T <1/2\pi R$, while the metric (\ref{204}) is dominant in the high temperature $T >1/2\pi R$. The transition between the metric (\ref{201}) and the metric (\ref{204}) occurs at a temperature of $T_{d} = 1/2\pi R \simeq 0.159/R$. This transition is first-order and corresponds to the confinement-deconfinement phase transition in the dual gauge theory 
side. 

In the deconfinement background, there are two kinds of configurations of D8-branes and $\overline{\text{D8}}$-branes as shown in Fig. 2. One is connected 
configuration and the other is disconnected configuration that the $\text{D8}$-branes and $\overline{\text{D8}}$-branes hang vertically from infinity down to 
the horizon. The disconnected configuration of the 
$\text{D8-}\overline{{\text{D8}}}$-branes indicates that the 
${\rm U}({\rm N}_{f})_{L} \times {\rm U}({\rm N}_{f})_{R}$ global chiral 
symmetry is restored in the dual gauge theory side. The transition between 
connected-disconnected configuration (chiral phase transition in the dual 
gauge theory side) is also first-order. We refer to the 
disconnected configuration and the connected configuration in the deconfinement background as ``parallel-embedding" of D8-branes and $\overline{\text{D8}}$-branes in the high-temperature phase and ``U-shaped embedding" of D8-branes and $\overline{\text{D8}}$-branes in the intermediate-temperature phase, respectively. The intermediate-temperature phase is realized when the confinement-deconfinement phase transition and the chiral phase transition does not occur simultaneously. 

\begin{figure}[H]
\centering
\hspace*{-15mm}
\begin{tabular}{cc}
\scalebox{0.6}{
\unitlength 0.1in
\begin{picture}( 52.1000, 24.1500)(  2.9000,-30.3000)
%
\special{pn 8}%
\special{ar 4300 1246 800 200  0.0000000 6.2831853}%
\put(43.0000,-8.4500){\makebox(0,0){\LARGE $t_{E}$}}%
%
\special{pn 8}%
\special{ar 2300 2650 800 200  6.2831853 6.2831853}%
\special{ar 2300 2650 800 200  0.0000000 3.1415927}%
%
\special{pn 8}%
\special{pa 1500 1250}%
\special{pa 1500 2650}%
\special{fp}%
\special{pa 3100 1250}%
\special{pa 3100 2650}%
\special{fp}%
\put(13.0000,-7.0000){\makebox(0,0){\LARGE $u$}}%
%
\special{pn 8}%
\special{ar 2300 1250 800 200  0.0000000 6.2831853}%
%
\special{pn 8}%
\special{ar 4300 1250 800 1600  6.2831853 6.2831853}%
\special{ar 4300 1250 800 1600  0.0000000 3.1415927}%
%
\special{pn 13}%
\special{ar 2300 1390 600 1200  6.2831853 6.2831853}%
\special{ar 2300 1390 600 1200  0.0000000 3.1415927}%
\put(11.0000,-28.6000){\makebox(0,0){\LARGE $u_{T}$}}%
\put(11.0000,-25.9000){\makebox(0,0){\LARGE $u_{0}$}}%
\put(23.0000,-8.5000){\makebox(0,0){\LARGE $\tau(=x_{4})$}}%
\put(23.0000,-22.0000){\makebox(0,0){\LARGE ${\rm D8}\text{-}\overline{\rm D8}$}}%
%
\special{pn 8}%
\special{pa 1300 3030}%
\special{pa 1300 830}%
\special{fp}%
\special{sh 1}%
\special{pa 1300 830}%
\special{pa 1280 898}%
\special{pa 1300 884}%
\special{pa 1320 898}%
\special{pa 1300 830}%
\special{fp}%
%
\special{pn 8}%
\special{pa 1300 2590}%
\special{pa 5500 2590}%
\special{dt 0.045}%
%
\special{pn 8}%
\special{ar 2300 1450 800 200  0.7853982 2.3561945}%
%
\special{pn 8}%
\special{pa 1752 1596}%
\special{pa 1734 1592}%
\special{fp}%
\special{sh 1}%
\special{pa 1734 1592}%
\special{pa 1794 1628}%
\special{pa 1786 1606}%
\special{pa 1804 1590}%
\special{pa 1734 1592}%
\special{fp}%
%
\special{pn 8}%
\special{pa 2848 1596}%
\special{pa 2866 1592}%
\special{fp}%
\special{sh 1}%
\special{pa 2866 1592}%
\special{pa 2796 1590}%
\special{pa 2816 1606}%
\special{pa 2808 1628}%
\special{pa 2866 1592}%
\special{fp}%
\put(23.0000,-18.0000){\makebox(0,0){\LARGE $L$}}%
%
\special{pn 8}%
\special{pa 2900 1250}%
\special{pa 3100 1250}%
\special{fp}%
\special{sh 1}%
\special{pa 3100 1250}%
\special{pa 3034 1230}%
\special{pa 3048 1250}%
\special{pa 3034 1270}%
\special{pa 3100 1250}%
\special{fp}%
\special{pa 2500 1250}%
\special{pa 2300 1250}%
\special{fp}%
\special{sh 1}%
\special{pa 2300 1250}%
\special{pa 2368 1270}%
\special{pa 2354 1250}%
\special{pa 2368 1230}%
\special{pa 2300 1250}%
\special{fp}%
\put(27.0000,-12.5000){\makebox(0,0){\LARGE $R$}}%
%
\special{pn 8}%
\special{pa 4900 1250}%
\special{pa 5100 1250}%
\special{fp}%
\special{sh 1}%
\special{pa 5100 1250}%
\special{pa 5034 1230}%
\special{pa 5048 1250}%
\special{pa 5034 1270}%
\special{pa 5100 1250}%
\special{fp}%
\special{pa 4500 1250}%
\special{pa 4300 1250}%
\special{fp}%
\special{sh 1}%
\special{pa 4300 1250}%
\special{pa 4368 1270}%
\special{pa 4354 1250}%
\special{pa 4368 1230}%
\special{pa 4300 1250}%
\special{fp}%
\put(47.0000,-12.5000){\makebox(0,0){\LARGE $R'$}}%
%
\special{pn 8}%
\special{ar 4300 2390 800 200  0.9600704 2.1815223}%
%
\special{pn 13}%
\special{pa 4860 2400}%
\special{pa 3870 1410}%
\special{dt 0.045}%
\special{pa 4830 2430}%
\special{pa 3800 1400}%
\special{dt 0.045}%
\special{pa 4810 2470}%
\special{pa 3720 1380}%
\special{dt 0.045}%
\special{pa 4790 2510}%
\special{pa 3630 1350}%
\special{dt 0.045}%
\special{pa 4760 2540}%
\special{pa 3500 1280}%
\special{dt 0.045}%
\special{pa 4720 2560}%
\special{pa 3510 1350}%
\special{dt 0.045}%
\special{pa 4670 2570}%
\special{pa 3510 1410}%
\special{dt 0.045}%
\special{pa 4610 2570}%
\special{pa 3510 1470}%
\special{dt 0.045}%
\special{pa 4560 2580}%
\special{pa 3510 1530}%
\special{dt 0.045}%
\special{pa 4500 2580}%
\special{pa 3520 1600}%
\special{dt 0.045}%
\special{pa 4440 2580}%
\special{pa 3530 1670}%
\special{dt 0.045}%
\special{pa 4390 2590}%
\special{pa 3540 1740}%
\special{dt 0.045}%
\special{pa 4330 2590}%
\special{pa 3550 1810}%
\special{dt 0.045}%
\special{pa 4270 2590}%
\special{pa 3570 1890}%
\special{dt 0.045}%
\special{pa 4210 2590}%
\special{pa 3590 1970}%
\special{dt 0.045}%
\special{pa 4140 2580}%
\special{pa 3610 2050}%
\special{dt 0.045}%
\special{pa 4080 2580}%
\special{pa 3640 2140}%
\special{dt 0.045}%
\special{pa 4020 2580}%
\special{pa 3670 2230}%
\special{dt 0.045}%
\special{pa 3950 2570}%
\special{pa 3720 2340}%
\special{dt 0.045}%
\special{pa 3880 2560}%
\special{pa 3780 2460}%
\special{dt 0.045}%
\special{pa 4880 2360}%
\special{pa 3940 1420}%
\special{dt 0.045}%
\special{pa 4890 2310}%
\special{pa 4010 1430}%
\special{dt 0.045}%
\special{pa 4910 2270}%
\special{pa 4070 1430}%
\special{dt 0.045}%
\special{pa 4930 2230}%
\special{pa 4140 1440}%
\special{dt 0.045}%
\special{pa 4950 2190}%
\special{pa 4200 1440}%
\special{dt 0.045}%
\special{pa 4960 2140}%
\special{pa 4260 1440}%
\special{dt 0.045}%
\special{pa 4980 2100}%
\special{pa 4320 1440}%
\special{dt 0.045}%
\special{pa 4990 2050}%
\special{pa 4380 1440}%
\special{dt 0.045}%
\special{pa 5000 2000}%
\special{pa 4440 1440}%
\special{dt 0.045}%
\special{pa 5010 1950}%
\special{pa 4490 1430}%
\special{dt 0.045}%
%
\special{pn 13}%
\special{pa 5030 1910}%
\special{pa 4550 1430}%
\special{dt 0.045}%
\special{pa 5040 1860}%
\special{pa 4610 1430}%
\special{dt 0.045}%
\special{pa 5050 1810}%
\special{pa 4660 1420}%
\special{dt 0.045}%
\special{pa 5050 1750}%
\special{pa 4710 1410}%
\special{dt 0.045}%
\special{pa 5060 1700}%
\special{pa 4760 1400}%
\special{dt 0.045}%
\special{pa 5070 1650}%
\special{pa 4810 1390}%
\special{dt 0.045}%
\special{pa 5080 1600}%
\special{pa 4860 1380}%
\special{dt 0.045}%
\special{pa 5080 1540}%
\special{pa 4910 1370}%
\special{dt 0.045}%
\special{pa 5090 1490}%
\special{pa 4960 1360}%
\special{dt 0.045}%
\special{pa 5090 1430}%
\special{pa 5000 1340}%
\special{dt 0.045}%
\special{pa 5090 1370}%
\special{pa 5040 1320}%
\special{dt 0.045}%
\special{pa 5100 1320}%
\special{pa 5070 1290}%
\special{dt 0.045}%
%
\special{pn 8}%
\special{pa 1300 2860}%
\special{pa 5500 2860}%
\special{dt 0.045}%
\end{picture}%
} & \scalebox{0.6}{
{\unitlength 0.1in%
\begin{picture}(48.2500,23.9500)(6.7500,-30.3000)%
%
\special{pn 8}%
\special{ar 4300 1245 800 200 0.0000000 6.2831853}%
\put(43.0000,-8.4500){\makebox(0,0){\LARGE $t_{E}$}}%
\put(13.0000,-7.0000){\makebox(0,0){\LARGE $u$}}%
%
\special{pn 8}%
\special{ar 2300 1250 800 200 0.0000000 6.2831853}%
%
\special{pn 8}%
\special{ar 4300 1250 800 1600 6.2831853 3.1415927}%
\put(10.8000,-28.5000){\makebox(0,0){\LARGE $u_{T}$}}%
\put(23.0000,-8.5000){\makebox(0,0){\LARGE $\tau(=x_{4})$}}%
\put(23.0000,-22.0000){\makebox(0,0){\LARGE ${\rm D8}\text{-}\overline{\rm D8}$}}%
%
\special{pn 8}%
\special{pa 1300 3030}%
\special{pa 1300 830}%
\special{fp}%
\special{sh 1}%
\special{pa 1300 830}%
\special{pa 1280 897}%
\special{pa 1300 883}%
\special{pa 1320 897}%
\special{pa 1300 830}%
\special{fp}%
%
\special{pn 8}%
\special{pa 1300 2850}%
\special{pa 5500 2850}%
\special{dt 0.045}%
%
\special{pn 8}%
\special{ar 2300 1450 800 200 0.7853982 2.3561945}%
%
\special{pn 8}%
\special{pa 1752 1596}%
\special{pa 1734 1591}%
\special{fp}%
\special{sh 1}%
\special{pa 1734 1591}%
\special{pa 1793 1628}%
\special{pa 1785 1605}%
\special{pa 1804 1590}%
\special{pa 1734 1591}%
\special{fp}%
%
\special{pn 8}%
\special{pa 2848 1596}%
\special{pa 2866 1591}%
\special{fp}%
\special{sh 1}%
\special{pa 2866 1591}%
\special{pa 2796 1590}%
\special{pa 2815 1605}%
\special{pa 2807 1628}%
\special{pa 2866 1591}%
\special{fp}%
\put(23.0000,-18.0000){\makebox(0,0){\LARGE $L$}}%
%
\special{pn 8}%
\special{pa 2900 1250}%
\special{pa 3100 1250}%
\special{fp}%
\special{sh 1}%
\special{pa 3100 1250}%
\special{pa 3033 1230}%
\special{pa 3047 1250}%
\special{pa 3033 1270}%
\special{pa 3100 1250}%
\special{fp}%
\special{pa 2500 1250}%
\special{pa 2300 1250}%
\special{fp}%
\special{sh 1}%
\special{pa 2300 1250}%
\special{pa 2367 1270}%
\special{pa 2353 1250}%
\special{pa 2367 1230}%
\special{pa 2300 1250}%
\special{fp}%
\put(27.0000,-12.5000){\makebox(0,0){\LARGE $R$}}%
%
\special{pn 8}%
\special{pa 4900 1250}%
\special{pa 5100 1250}%
\special{fp}%
\special{sh 1}%
\special{pa 5100 1250}%
\special{pa 5033 1230}%
\special{pa 5047 1250}%
\special{pa 5033 1270}%
\special{pa 5100 1250}%
\special{fp}%
\special{pa 4500 1250}%
\special{pa 4300 1250}%
\special{fp}%
\special{sh 1}%
\special{pa 4300 1250}%
\special{pa 4367 1270}%
\special{pa 4353 1250}%
\special{pa 4367 1230}%
\special{pa 4300 1250}%
\special{fp}%
\put(47.0000,-12.5000){\makebox(0,0){\LARGE $R'$}}%
%
\special{pn 15}%
\special{pa 1720 1390}%
\special{pa 1720 2790}%
\special{fp}%
%
\special{pn 15}%
\special{pa 2900 1380}%
\special{pa 2900 2780}%
\special{fp}%
%
\special{pn 8}%
\special{ar 2300 2650 800 200 6.2831853 3.1415927}%
%
\special{pn 8}%
\special{pa 1510 1250}%
\special{pa 1510 2650}%
\special{fp}%
\special{pa 3110 1250}%
\special{pa 3110 2650}%
\special{fp}%
%
\special{pn 15}%
\special{pa 4690 2650}%
\special{pa 3510 1470}%
\special{dt 0.045}%
\special{pa 4710 2610}%
\special{pa 3510 1410}%
\special{dt 0.045}%
\special{pa 4740 2580}%
\special{pa 3510 1350}%
\special{dt 0.045}%
\special{pa 4770 2550}%
\special{pa 3500 1280}%
\special{dt 0.045}%
\special{pa 4790 2510}%
\special{pa 3630 1350}%
\special{dt 0.045}%
\special{pa 4810 2470}%
\special{pa 3720 1380}%
\special{dt 0.045}%
\special{pa 4830 2430}%
\special{pa 3800 1400}%
\special{dt 0.045}%
\special{pa 4860 2400}%
\special{pa 3870 1410}%
\special{dt 0.045}%
\special{pa 4880 2360}%
\special{pa 3940 1420}%
\special{dt 0.045}%
\special{pa 4890 2310}%
\special{pa 4010 1430}%
\special{dt 0.045}%
\special{pa 4910 2270}%
\special{pa 4070 1430}%
\special{dt 0.045}%
\special{pa 4930 2230}%
\special{pa 4140 1440}%
\special{dt 0.045}%
\special{pa 4950 2190}%
\special{pa 4200 1440}%
\special{dt 0.045}%
\special{pa 4960 2140}%
\special{pa 4260 1440}%
\special{dt 0.045}%
\special{pa 4980 2100}%
\special{pa 4320 1440}%
\special{dt 0.045}%
\special{pa 4990 2050}%
\special{pa 4380 1440}%
\special{dt 0.045}%
\special{pa 5000 2000}%
\special{pa 4440 1440}%
\special{dt 0.045}%
\special{pa 5010 1950}%
\special{pa 4490 1430}%
\special{dt 0.045}%
\special{pa 5030 1910}%
\special{pa 4550 1430}%
\special{dt 0.045}%
\special{pa 5040 1860}%
\special{pa 4610 1430}%
\special{dt 0.045}%
\special{pa 5050 1810}%
\special{pa 4660 1420}%
\special{dt 0.045}%
\special{pa 5050 1750}%
\special{pa 4710 1410}%
\special{dt 0.045}%
\special{pa 5060 1700}%
\special{pa 4760 1400}%
\special{dt 0.045}%
\special{pa 5070 1650}%
\special{pa 4810 1390}%
\special{dt 0.045}%
\special{pa 5080 1600}%
\special{pa 4860 1380}%
\special{dt 0.045}%
\special{pa 5080 1540}%
\special{pa 4910 1370}%
\special{dt 0.045}%
\special{pa 5090 1490}%
\special{pa 4960 1360}%
\special{dt 0.045}%
\special{pa 5090 1430}%
\special{pa 5000 1340}%
\special{dt 0.045}%
\special{pa 5090 1370}%
\special{pa 5040 1320}%
\special{dt 0.045}%
\special{pa 5100 1320}%
\special{pa 5070 1290}%
\special{dt 0.045}%
%
\special{pn 15}%
\special{pa 4660 2680}%
\special{pa 3510 1530}%
\special{dt 0.045}%
\special{pa 4630 2710}%
\special{pa 3520 1600}%
\special{dt 0.045}%
\special{pa 4590 2730}%
\special{pa 3530 1670}%
\special{dt 0.045}%
\special{pa 4560 2760}%
\special{pa 3540 1740}%
\special{dt 0.045}%
\special{pa 4520 2780}%
\special{pa 3550 1810}%
\special{dt 0.045}%
\special{pa 4480 2800}%
\special{pa 3570 1890}%
\special{dt 0.045}%
\special{pa 4440 2820}%
\special{pa 3590 1970}%
\special{dt 0.045}%
\special{pa 4400 2840}%
\special{pa 3610 2050}%
\special{dt 0.045}%
\special{pa 4350 2850}%
\special{pa 3640 2140}%
\special{dt 0.045}%
\special{pa 4290 2850}%
\special{pa 3670 2230}%
\special{dt 0.045}%
\special{pa 4220 2840}%
\special{pa 3720 2340}%
\special{dt 0.045}%
\special{pa 4130 2810}%
\special{pa 3780 2460}%
\special{dt 0.045}%
\end{picture}}%
} \\
\end{tabular}
\caption{The $\text{D8-}\overline{\text{D8}}$-branes configurations at high temperature.}
\label{HT}
\end{figure}
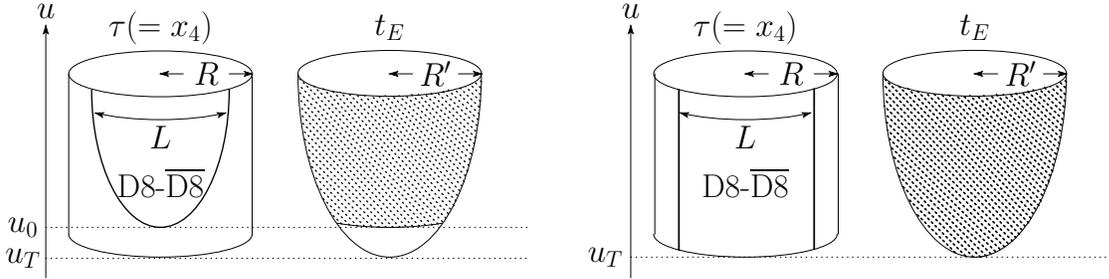

As mentioned above, the classical configuration of D8-branes and $\overline{\text{D8}}$-branes exhibits the flavour physics in the dual gauge theory side. The configuration can be analysed by the solution of the equation of motion for the 
D8-branes. Substituting the determinant of the induced metric in the deconfining background and the dilaton into the Dirac--Born--Infeld(DBI) action, we obtain the effective action for the D8-branes:
\begin{align}
\label{206}
S^{\rm D8}_{\rm DBI}
&=2N_{f}T_{8}\int d^{9}x\,e^{-\phi}\sqrt{\det\,(g_{MN})} \nonumber \\
&=\dfrac{2N_{f}T_{8}V_{4}}{g_{s}}\int d^{4}x\,du\;u^{4} \sqrt{
f_{T}(u)\tau'(u)^{2} + \dfrac{R_{\rm D4}^{3}}{u^{3}}} \;, 
\end{align}
where $T_{8}$ is the tension of the D8-brane and the prime of $\tau$ denotes differentiation with respect to $u$. The constant of motion associated with $\tau$, denoted by $p$, has the following form, 
\begin{align}
\label{207}
\dfrac{u^{4}f_{T}(u)\tau'(u)}{\sqrt{f_{T}(u)\tau'(u)^{2}+\frac{R_{\rm D4}^{3}}{u^{3}}}} 
= p =u_{0}^{4}\sqrt{f_{T}(u_{0})}\;,
\end{align}
where we assumed that there is a point $u_{0}$ that satisfies the condition $\lim_{u \to u_{0}}\tau'(u) \to \infty$. The solution to the equation of motion for $\tau(u)$ is found to be
\begin{align}
\label{208}
\tau'(u) &=\sqrt{\dfrac{R_{\rm D4}^{3}}{u^{3}f_{T}(u)}} 
\left[\dfrac{u^{8}f_{T}(u)}
{u_{0}^{8}f_{T}(u_{0})}-1 \right]^{-1/2} \;,
\end{align}
by using (\ref{207}). This solution corresponds to the U-shaped embedding of D8-branes and $\overline{\text{D8}}$-branes. There is another solution to the equation of motion for $\tau(u)$ in the deconfinement background. This solution is simply given by $\tau'(u)=0\;(\tau(u)\,\text{is a constant})$ and corresponds to the parallel embedding of D8-branes and $\overline{\text{D8}}$-branes. 

The asymptotic D8-branes and $\overline{\text{D8}}$-branes distance can be obtained by integrating (\ref{208}) with respect to $u$: 
\begin{align}
\label{209}
L = \int d\tau = 2\int^{\infty}_{u_{0}}du\, \tau'(u)\;.
\end{align}
The asymptotic distance $L_{\chi}$ and the temperature $T_{\chi}$ at the chiral symmetry phase transition can be related as $L_{\chi}T_{\chi} \simeq 0.154$. For $LT<0.154$ the U-shaped embedding dominates and chiral symmetry is broken. On the other hand, for $LT>0.154$ the parallel embedding dominates and chiral symmetry is restored. When $T_{\chi}$ is higher than $T_{d}$, namely small $L/R\,(<0.97)$, the dual gauge theory is deconfined but with a broken chiral symmetry \cite{ASY}.

The constant of motion $p$ remains a finite value that is given by (\ref{207}) in the U-shaped embedding with a broken chiral symmetry and vanishes in the parallel embedding with a restored chiral symmetry. In this sense, we can regard $p$ as an order parameter for the chiral transition in the deconfined phase. This first order phase transition behavior can be analysed from the dependence of the asymptotic distance $L$ on $p$ \cite{BLL}.

The holographic dual description of the noncommutative gauge theories was introduced in \cite{HI, MR, AOSJ}. In accordance with the formulation of 
\cite{HI, MR, AOSJ}, we attempt to construct the gravity dual of the 
noncommutative QCD whose chiral symmetry is spontaneously broken by deforming 
the holographic QCD model. Let us consider the D4-branes solution compactified 
on a circle in the $\tau$-direction. T-dualizing it along $x^{3}$ produces a 
D3-branes delocalized along $x^{3}$. After rotating the D3-branes along the 
($x^{2},\;x^{3}$) plane, we T-dualize back on $x^{3}$. This procedure yields 
the solution with a $B_{23}$ fields along the $x^{2}$ and $x^{3}$ directions. 
The solution in the low temperature takes the form 
\begin{align}
ds^{2} &=\left(\frac{u}{R_{\rm D4}}\right)^{3/2}
\Bigl((dt_{E})^{2}+(dx^{1})^{2}
+ h\{(dx^{2})^{2}+(dx^{3})^{2}\} + f_{K}(u)\,d\tau^{2} \Bigr) 
\nonumber \\
&+ \left(\frac{R_{\rm D4}}{u}\right)^{3/2}
\left(\frac{du^{2}}{f_{K}(u)}+u^{2}d\Omega_{4}^{2} \right)\,,
\label{210}
\end{align}
where $h(u)=\dfrac{1}{1+\theta^{3}u^{3}}$ and $\theta$ denotes the 
noncommutativity parameter with dimension of $[{\rm length}]^{-1}$. This 
solution with $\theta \neq 0$ is dual to a gauge theory in which the 
coordinates $x^{2}$ and $x^{3}$ do not commute. It is obvious that this 
solution reduces to the solution (\ref{201}) with Euclidean signature at 
$\theta =0$. In the deconfined phase, the solution (\ref{204}) changes to 
\begin{align}
ds^{2} &=\left(\frac{u}{R_{\rm D4}}\right)^{3/2}
\Bigl(f_{T}(u)(dx_{E})^{2}+(dx^{1})^{2}
+ h\{(dx^{2})^{2}+(dx^{3})^{2}\} + d\tau^{2} \Bigr) 
\nonumber \\
&+ \left(\frac{R_{\rm D4}}{u}\right)^{3/2}
\left(\frac{du^{2}}{f_{T}(u)}+u^{2}d\Omega_{4}^{2} \right)\,. 
\label{211}
\end{align}
The solution has the same form as the one in the confined phase (\ref{210}), but with the role of the $\tau$ and $t_{E}$ directions exchanged. 

The effective action of probe D8-branes is given by the DBI action with the Wess--Zumino(WZ) term: 
\begin{align}
\label{212}
S^{\rm D8} &=S^{\rm D8}_{\rm DBI} + S^{\rm D8}_{\rm WZ}\,, \\ 
& S^{\rm D8}_{\rm DBI}
=T_{8}\int d^{9}x e^{-\phi}{\rm Tr}\sqrt{\det (g_{MN} + B_{MN} 
+ 2\pi \alpha'F_{MN})}\,, 
\nonumber \\
& S^{\rm D8}_{\rm WZ}
=\mu_{8}\int_{\rm D8}\,C_{3} \wedge e^{(\,\widetilde{B}+2\pi l_{s}{}^{2}F\,)}\,, \nonumber 
\end{align}
where $\mu_{8}$ is the D8-brane charge. The dilaton field $\phi$ and the antisymmetric tensor field $\widetilde{B}=B_{MN}dx^{M}dx^{N}$ have the following form: 
\begin{align}
\label{213}
&e^{2\phi} = g_{s}{}^{2}h(u)\left(\dfrac{u}{R_{\rm D4}} \right)^{3/2} \;, 
\\[3mm] 
\label{214}
&B_{MN}(u)=\left\{
\begin{array}{ll}
\theta^{3/2}\dfrac{u^{3}}{R_{\rm D4}^{3/2}}h(u) & (M=2,\;N=3) \\
0 & (\text{others})
\end{array} \right.\;.
\end{align}
We notice that the dependence of DBI action on the noncommutativity parameter $\theta$ is canceled by the dilaton and the antisymmetric tensor field. The cancellation of the noncommutativity parameter dependence in the DBI action also takes place in the effective action of the probe D7-brane \cite{APR}. Adding the 
WZ-term to the DBI action, we find the dependence on the noncommutativity parameter in the effective action of the D8-branes. Hereafter the parameter $R_{\rm D4}$ is fixed to unity, $R_{\rm D4}=(\pi g_{s}N_{c})^{1/3}l_{s}=1$, for simplicity.

%
%
\section{Electric Field} 
\setcounter{equation}{0}
\addtocounter{enumi}{1}

We investigate the response of the noncommutative deformation of holographic QCD at finite temperature to an external electric field $E$, by turning on an appropriate background value for the abelian gauge field component of the unbroken $U(N_{f})_{V}$ gauge field in the 8-brane world volume. 

We make an ansatz
\begin{align}
\label{301}
2\pi \alpha'A_{0}=\mu \;, \quad 
2\pi \alpha'A_{1}(t_{E},\;u)=-iet_{E}+a_{1}(u)\;,
\end{align}
where $\mu$ and $e$ are constants. 

\subsection{Deconfinement phase} 

We first consider the deconfining background, which dominates at high temperature $T >1/2\pi R$. The induced metric on the probe D8-brane is 
\begin{align}
\label{302}
ds^{2}&=u^{3/2}\bigl(f_{T}(u)dt_{E}^{2}+(dx^{1})^{2}
+h(u)\{(dx^{2})^{2}+(dx^{3})^{2}\} \bigr) \nonumber \\
&+\left[u^{3/2}(\tau'(u))^{2}+u^{-3/2}f_{T}(u)^{-1} \right]du^{2} 
+ u^{1/2}d\Omega_{4}^{2}\;,
\end{align}
where the temperature $T$ is related to the parameter $u_{T}$ as $u_{T}=\left(
16\pi^{2}/9 \right)T^{2}$. The DBI action with the WZ term takes the form 
\begin{align}
\label{303}
S^{\rm D8} &=S^{\rm D8}_{\rm DBI} + S^{\rm D8}_{\rm WZ} \nonumber \\
&={\cal N}\int d^{4}x du \; \biggl[ u^{4} \sqrt{
\left(f_{T}(u)\tau'(u)^{2} + \dfrac{1}{u^{3}} \right)
\left(1 + \dfrac{e^{2}}{u^{3}f_{T}(u)} \right)
+ \dfrac{f_{T}(u)(a'_{1}(u))^{2}}{u^{3}} } \Biggr. \nonumber \\
& \hspace{30mm} \biggl. - 3\mu \theta^{3/2}u^{3}h(u)a_{1}'(u) \biggr] \;,
\end{align}
where ${\cal N}=\dfrac{2N_{f}N_{c}}{3(2\pi)^{5}(\alpha')^{3}}$ and the prime of $a_{1}$ denotes differentiation with respect to $u$. The baryon number current $j_{eT}$ associated with the field $a_{1}$ is expressed as 
\begin{align}
\label{304}
j_{eT}=\dfrac{uf_{T}(u)a'_{1}(u)}{\sqrt{
\left(f_{T}(u)\tau'(u)^{2} + \dfrac{1}{u^{3}} \right)
\left(1 + \dfrac{e^{2}}{u^{3}f_{T}(u)} \right)
+ \dfrac{f_{T}(u)(a'_{1}(u))^{2}}{u^{3}} }}
- 3\mu \theta^{3/2}u^{3}h(u) \;.
\end{align}
The DBI action with the WZ term can be written in terms of the baryon number current as
\begin{align}
\label{305}
S^{\rm D8} &={\cal N}\int d^{4}x du \, u^{4}
\left\{1 - \dfrac{3\mu \theta^{3/2}h(u)\widetilde{j}_{eT}}{u^{2}f_{T}(u)} 
\right\}
\sqrt{\dfrac{\left(f_{T}(u)\tau'(u)^{2} + \dfrac{1}{u^{3}} \right)
\left(f_{T}(u) - \dfrac{e^{2}}{u^{3}} \right)}
{\left(f_{T}(u) - \dfrac{\widetilde{j}_{eT}^{2}}{u^{2}} \right)}} \;,
\end{align}
where $\widetilde{j}_{eT}=j_{eT}+3\mu\theta^{3/2}u^{3}h(u)$. Consider first a U-shaped embedding with a vanishing current $j_{eT}=0$. The corresponding action is given by
\begin{align}
\label{306}
S^{\rm D8} &={\cal N}\int d^{4}x du \;
\dfrac{u^{4}}{f_{T}(u)}
\sqrt{
\left(f_{T}(u)\tau'(u)^{2} + \dfrac{1}{u^{3}} \right)
\left(f_{T}(u) - \dfrac{e^{2}}{u^{3}} \right)
\Bigl(f_{T}(u) - 9\mu^{2}\theta^{3}uh(u)^{2} \Bigr)} \;.
\end{align}
The equation of motion for $\tau(u)$ is 
\begin{align}
\label{307}
\dfrac{d}{du}
\left\{u^{4}\tau'(u)\sqrt{\dfrac{\left(f_{T}(u)-\dfrac{e^{2}}{u^{3}}\right)
\Bigl(f_{T}(u)-9\mu^{2}\theta^{3}uh(u)^{2} \Bigr)}
{\left(f_{T}(u)(\tau'(u))^{2}+\dfrac{1}{u^{3}}\right)}}\right\}=0 \;.
\end{align}
$\tau'(u)$ satisfies the condition in U-shaped embedding configuration: $\tau' \to \infty$ for $u \to u_{0}$. In the limit $u \to u_{0}$, we have the constant of the motion associated with $\tau(u)$ as
\begin{align}
\label{308}
p_{1}=u_{0}^{4}\sqrt{\left(1-\dfrac{e^{2}}{u_{0}^{3}f_{T}(u_{0})}\right)
\Bigl(f_{T}(u_{0})-9\mu^{2}\theta^{3}u_{0}h(u_{0})^{2} \Bigr)}\;.
\end{align}
The solution of the equation of motion for $\tau(u)$ is
\begin{align}
\label{309}
\tau'(u) &=\dfrac{1}{u^{3/2}\sqrt{f_{T}(u)}}
\left[\dfrac{u^{8}\left(1-\dfrac{e^{2}}{u^{3}f_{T}(u)}\right)
\Bigl(f_{T}(u)-9\mu^{2}\theta^{3}uh(u)^{2}\Bigr)}
{u_{0}^{8}\left(1-\dfrac{e^{2}}{u_{0}^{3}f_{T}(u_{0})}\right)
\Bigl(f_{T}(u_{0})-9\mu^{2}\theta^{3}u_{0}h(u_{0})^{2}\Bigr)}-1 \right]^{-1/2} \;.
\end{align}
The reality conditions of the constant in (\ref{308}) restricts the 
parameters $e$ and $\theta$ as
\begin{align}
\label{310}
& e^{2} \leq u_{0}^{3}-u_{T}^{3} \;, 
\qquad \theta^{3} \leq K_{1-}\,,\; \theta^{3} \geq K_{1+} \;, \\
& \qquad K_{1\pm} \equiv \dfrac{(9\mu^{2}-2u_{0}^{2}f_{T}(u_{0}))
\pm 3\mu\sqrt{9\mu^{2}-4u_{0}^{2}f_{T}(u_{0})}}{2u_{0}^{5}f_{T}(u_{0})}\;.
\nonumber 
\end{align}
In the U-shaped embedding, the corresponding (on-shell) action is obtained by substituting (\ref{309}) into (\ref{306}):
\begin{align}
\label{311}
S^{\rm D8}_{\rm U} &={\cal N}\int d^{4}x du \; \dfrac{u^{5/2}}{f_{T}(u)} 
\sqrt{\left(f_{T}(u) - \dfrac{e^{2}}{u^{3}} \right)
\Bigl(f_{T}(u)- 9\mu^{2}\theta^{3}uh(u)^{2} \Bigr) } 
\nonumber \\
& \qquad \times \left[ 1- \dfrac{u_{0}^{8}
\left(1 - \dfrac{e^{2}}{u_{0}^{3}f_{T}(u_{0})} \right)
\Bigl(f_{T}(u_{0}) - 9\mu^{2}\theta^{3}uh(u_{0})^{2} \Bigr)}
{u^{8}\left(1 - \dfrac{e^{2}}{u^{3}f_{T}(u)} \right)
\Bigl(f_{T}(u) - 9\mu^{2}\theta^{3}uh(u)^{2} \Bigr)} 
\right]^{-1/2} \;.
\end{align}
In the parallel embedding with $\tau'(u)=0$, the action becomes
\begin{align}
\label{312}
S^{\rm D8}_{\rm ||} &={\cal N}\int d^{4}x du \; \dfrac{u^{5/2}}{f_{T}(u)} 
\left\{f_{T}(u)- \dfrac{3\mu\theta^{3/2}h(u) \widetilde{j}_{eT}}{u^{2}} \right\}
\sqrt{ \dfrac{f_{T}(u) - \dfrac{e^{2}}{u^{3}}}
{f_{T}(u) - \dfrac{\widetilde{j}_{eT}^{2}}{u^{5}}} } \;.
\end{align}
The numerator of fraction in the square root $f_{T}(u) - e^{2}/u^{3}$ is 
negative for $u^{3}<u_{T}^{3}+e^{2}$, which is always in the range of interaction. The only way to ensure a real action in this case is for the denominator in the same square root to become negative at the same $u$. This requires a nonvanishing current that is given by 
\begin{align}
\label{313}
j_{eT}=e(u_{T}^{3}+e^{2})^{1/3}
-\dfrac{3\mu\theta^{3/2}(u_{T}^{3}+e^{2})}{1+\theta^{3}(u_{T}^{3}+e^{2})} \;.
\end{align}
This current depends on the noncommutativity parameter $\theta$. The parallel embedding therefore describes a chiral-symmetric conducting phase in the gauge theory, and the conductivity is given by 
\begin{align}
\label{314}
\sigma_{e} &=\dfrac{(2\pi\alpha')^{2}{\cal N}}{V_{4}}\dfrac{j_{eT}}{e} 
\nonumber \\
&=\dfrac{N_{f}N_{c}\lambda T^{2}}{27\pi}
\left[ (1+\widetilde{e}^{2})^{1/3} 
- \dfrac{3\widetilde{\mu}\widetilde{\theta}^{3/2}(1+\widetilde{e}^{2})}
{\widetilde{e}\{1+\widetilde{\theta}^{3}(1+\widetilde{e}^{2})\}} \right]
\;,
\end{align}
where $\widetilde{e} \equiv e/u_{T}^{3/2}$, $\widetilde{\theta} \equiv u_{T}\theta$ and $\widetilde{\mu} \equiv \mu/u_{T}$ are dimensionless parameters. The 
conductivity depends on the noncommutativity parameter and becomes the ordinary one in the limit of $\theta \to 0$ \cite{BLL}. 

If the parallel embedding corresponds to the state of thermodynamic equilibrium, we can determined which of the two possible configuration is preferred by comparing the electric free energies of the two configurations \cite{ASY}. However, the parallel embedding corresponds to the conducting phase, which is not in thermodynamic equilibrium. There is a steady state current of quarks and anti-quarks. 
Although the dissipated energy could be negligible, the kinetic energy of the current carriers should be taken into consideration. 

\begin{figure}[H]
\centering
\includegraphics[width=40mm]{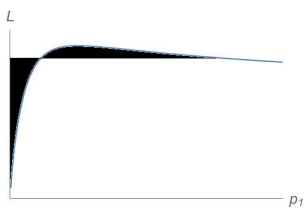}
\caption{Illustration of the Maxwell equal area construction}
\label{MC1}
\end{figure}

\begin{figure}[H]
\begin{minipage}{0.5\hsize}
\centering
\includegraphics[width=60mm]{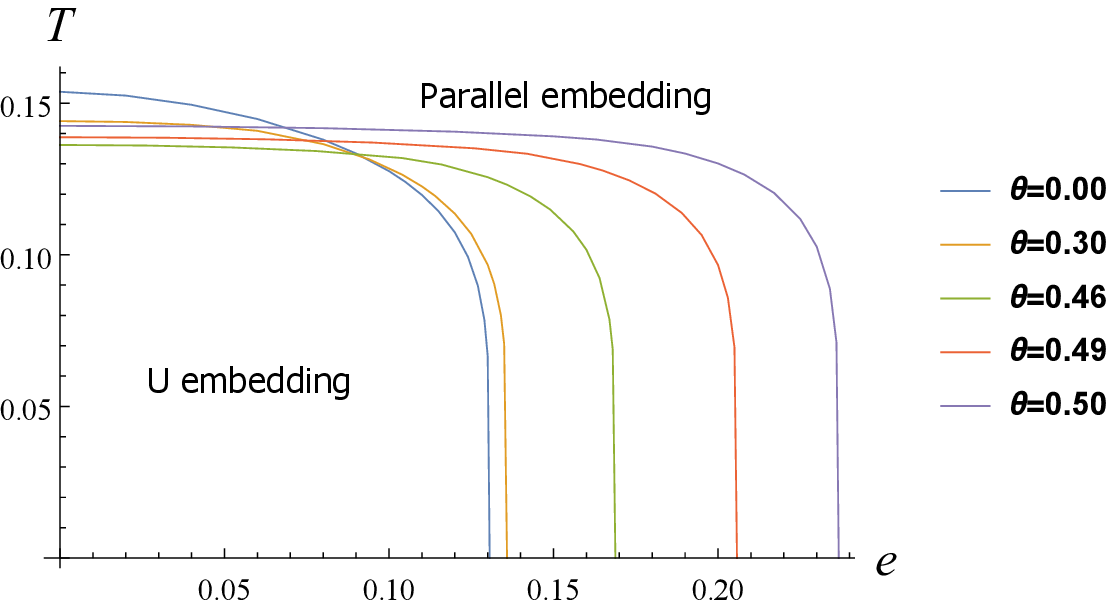} \\
$0.00 \leq \theta \leq 0.50$
\end{minipage}
\begin{minipage}{0.5\hsize}
\centering
\includegraphics[width=60mm]{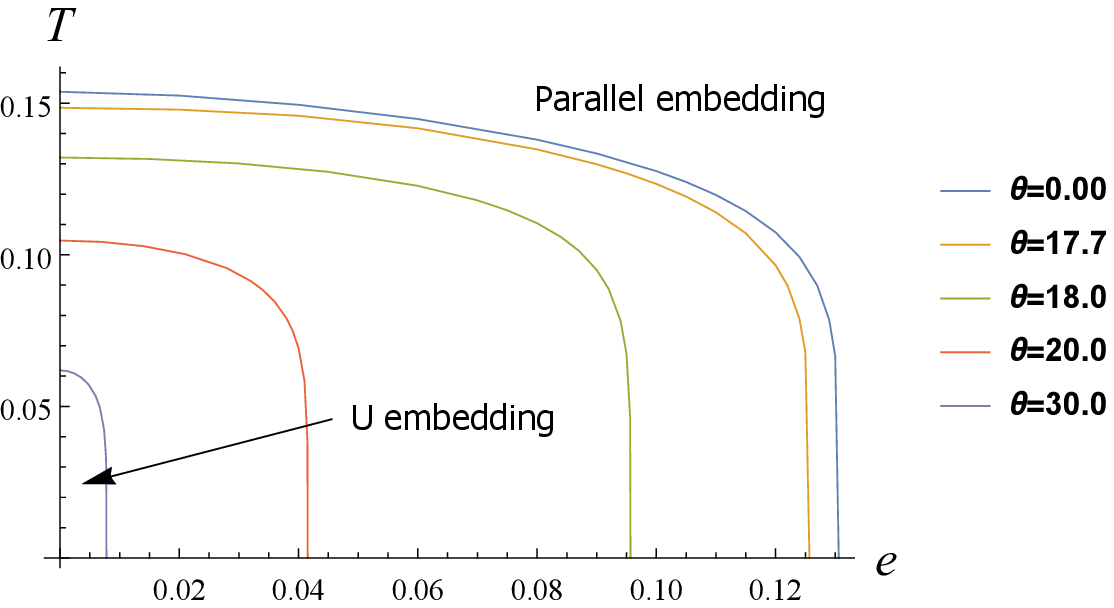} \\
$17.7 \leq \theta$
\end{minipage}
\caption{Phase diagram at finite temperature and electric field in the deconfining background. ($L=1$, $\mu=1$)}
\label{ETDC}
\end{figure}

In order to determine the transition temperature and transition electric field strength, we employ the Maxwell equal area construction method in the $L$-$p$ diagram. The dependence of $p_{1}$ on $L$ can be determined numerically from (\ref{308}) and (\ref{309})(with (\ref{209})) in the U-embedding configuration. The phase transition occurs when two regions enclosed by the $L$-$p$ curve and the horizontal $L$ line (and $L$-axis) are equal as shown by Fig.\ref{MC1}. 
We can determine the transition temperature $T_{c}$ and the transition electric field strength $e_{c}$ by seeking for various of $T$ and $e$ to satisfy the Maxwell equal area law and then construct the phase diagram in the $(T,\;e)$ plane with fixed $L$ and $\theta$. The phase diagram at nonzero temperature, background electric field and noncommutativity parameter in the deconfining phase is shown in Fig.\ref{ETDC}. At zero electric field and zero noncommutativity parameter, the transition temperature reduces to the one of chiral symmetry breaking restoration \cite{ASY}. \\

The global behavior of the phase diagrams has not significantly changed  even at finite noncommutativity parameter $\theta$, that is, the transition temperature $T_{c}$ decreases as the transition electric field strength $e_{c}$ increases even at finite noncommutativity parameter $\theta$. Although $T_{c}$ at $e_{c}=0$ is hardly changed, $e_{c}$ at $T_{c}=0$ increases with an increase in $\theta$ in the range of $0 \leq \theta \leq 0.50$. Both $T_{c}$ at $e_{c}=0$ and $e_{c}$ at $T_{c}=0$ decrease with decreasing $\theta$ in the range of $17.7 \leq \theta$. As $\theta$ approaches infinity, both $T_{c}$ and $e_{c}$ turn back to them at zero noncommutativity parameter. The reality condition is not satisfied in the range of $0.50 < \theta < 17.7$. 

\subsection{Confinement phase} 

We next consider the confining background, which dominates at low temperature $T < 1/2\pi R$. The induced metric on the probe D8-brane is 
\begin{align}
\label{315}
ds^{2}&=u^{3/2}\bigl(dt_{E}^{2}+(dx^{1})^{2}
+h(u)\{(dx^{2})^{2}+(dx^{3})^{2}\} \bigr) \nonumber \\
&+\left[u^{3/2}f_{K}(u)(\tau'(u))^{2}+u^{-3/2}f_{K}(u)^{-1} \right]du^{2} 
+ u^{1/2}d\Omega_{4}^{2}\;.
\end{align}
The total effective action for the D8-branes is given by 
\begin{align}
\label{316}
S^{\rm D8} &=S^{\rm D8}_{\rm DBI} + S^{\rm D8}_{\rm WZ} \nonumber \\
&={\cal N}\int d^{4}x du \; \biggl[ u^{4} \sqrt{
\left(f_{K}(u)\tau'(u)^{2} + \dfrac{1}{u^{3}f_{K}(u)} \right)
\left(1 - \dfrac{e^{2}}{u^{3}} \right)
+ \dfrac{(a'_{1}(u))^{2}}{u^{3}} } \Biggr. \nonumber \\
& \hspace{30mm} \biggl. - 3\mu \theta^{3/2}u^{3}h(u)a_{1}'(u) \biggr] 
\nonumber \\
& ={\cal N}\int d^{4}x du \; u^{4} 
\left\{ 1-\dfrac{3\mu \theta^{3/2}h(u)\widetilde{j}_{eK}}{u^{2}} \right\}
\sqrt{\dfrac{\left(f_{K}(u)\tau'(u)^{2} + \dfrac{1}{u^{3}f_{K}(u)} \right)
\left(1 - \dfrac{e^{2}}{u^{3}} \right)}
{\left(1 - \dfrac{\widetilde{j}_{eK}^{2}}{u^{2}} \right)}} \;,
\end{align}
where $\widetilde{j}_{eK}=j_{eK}+3\mu\theta^{3/2}u^{3}h(u)$ with 
\begin{align}
\label{317}
j_{eK}=\dfrac{ua'_{1}(u)}{\sqrt{
\left(f_{K}(u)\tau'(u)^{2} + \dfrac{1}{u^{3}f_{K}(u)} \right)
\left(1 - \dfrac{e^{2}}{u^{3}} \right)
+ \dfrac{(a'_{1}(u))^{2}}{u^{3}} }}
- 3\mu \theta^{3/2}u^{3}h(u) \;.
\end{align}
The solution of the equation of motion for $\tau(u)$ in the U-embedding (with the vanishing $j_{eK}$) is given by
\begin{align}
\label{318}
\tau'(u) &=\dfrac{1}{u^{3/2}f_{K}(u)} 
\left[\dfrac{u^{8}f_{K}(u)\left(1-\dfrac{e^{2}}{u^{3}}\right)
\Bigl(1-9\mu^{2}\theta^{3}uh(u)^{2}\Bigr)}
{u_{0}^{8}f_{K}(u_{0})\left(1-\dfrac{e^{2}}{u_{0}^{3}}\right)
\Bigl(1-9\mu^{2}\theta^{3}u_{0}h(u_{0})^{2}\Bigr)}-1 \right]^{-1/2} \;,
\end{align}
and the constant of motion for $\tau(u)$ is 
\begin{align}
\label{319}
p_{2}=u_{0}^{4}\sqrt{f_{K}(u_{0})\left(1-\dfrac{e^{2}}{u_{0}^{3}}\right)
\Bigl(1-9\mu^{2}\theta^{3}u_{0}h(u_{0})^{2} \Bigr)}\;.
\end{align}
In the same way as the deconfinement phase, the reality conditions of the constant in (\ref{319}) restricts the parameters $e$ and $\theta$, 
\begin{align}
\label{320}
& e^{2} \leq u_{0}^{3} \;, 
\qquad \theta^{3} \leq K_{2-}\,,\; \theta^{3} \geq K_{2+} \;, \\
& \qquad K_{2\pm} \equiv \dfrac{(9\mu^{2}-2u_{0}^{2})
\pm 3\mu\sqrt{9\mu^{2}-4u_{0}^{2}}}{2u_{0}^{5}}\;.
\nonumber 
\end{align}

\begin{figure}[H]
\begin{minipage}{0.5\hsize}
\centering
\includegraphics[width=50mm]{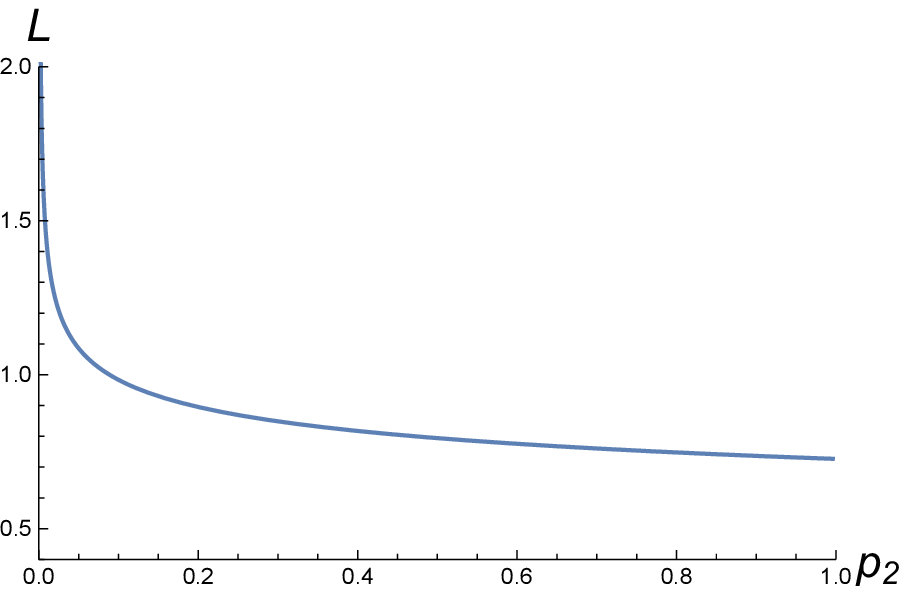} \\
$e^2 < u_{K}^{3}$
\end{minipage}
\begin{minipage}{0.5\hsize}
\centering
\includegraphics[width=50mm]{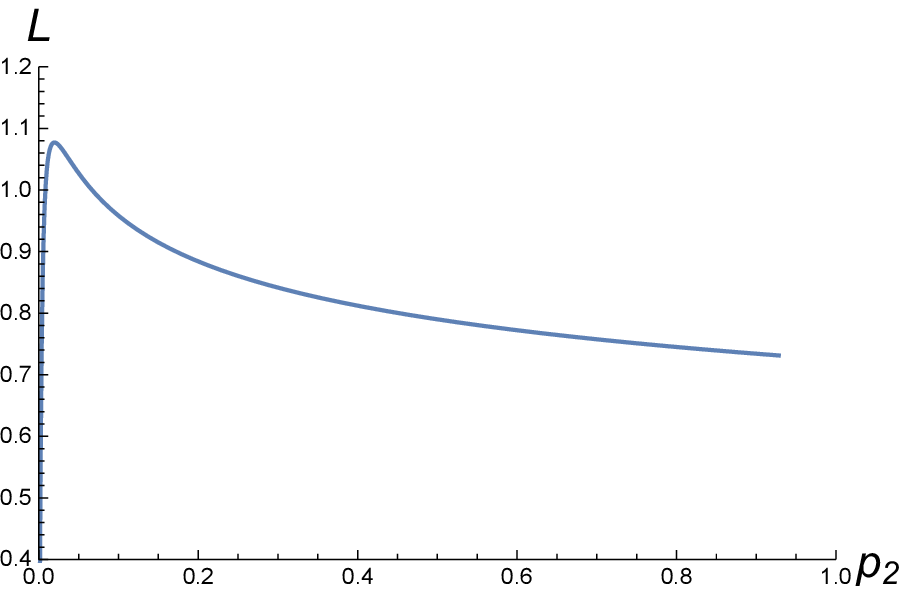} \\
$e^2 > u_{K}^{3}$
\end{minipage}
\caption{Dependence of the asymptotic $\text{D8-}\overline{\text{D8}}$ distance $L$ on $p_{2}$}
\label{LCC}
\end{figure}

\begin{figure}[H]
\centering
\scalebox{0.6}{
{\unitlength 0.1in%
\begin{picture}(21.6700,23.9500)(10.0000,-30.3000)%
\put(13.0000,-7.0000){\makebox(0,0){\LARGE $u$}}%
%
\special{pn 8}%
\special{ar 2300 1250 800 200 0.0000000 6.2831853}%
%
\special{pn 8}%
\special{ar 2300 1250 800 1600 6.2831853 3.1415927}%
\put(23.0000,-8.5000){\makebox(0,0){\LARGE $x_{4}$}}%
\put(23.0000,-22.0000){\makebox(0,0){\LARGE ${\rm D8}\text{-}\overline{\rm D8}$}}%
%
\special{pn 8}%
\special{pa 1300 3030}%
\special{pa 1300 830}%
\special{fp}%
\special{sh 1}%
\special{pa 1300 830}%
\special{pa 1280 897}%
\special{pa 1300 883}%
\special{pa 1320 897}%
\special{pa 1300 830}%
\special{fp}%
%
\special{pn 8}%
\special{ar 2287 1450 782 200 0.7853982 2.3561945}%
%
\special{pn 8}%
\special{pa 1752 1596}%
\special{pa 1734 1591}%
\special{fp}%
\special{sh 1}%
\special{pa 1734 1591}%
\special{pa 1793 1628}%
\special{pa 1785 1605}%
\special{pa 1804 1590}%
\special{pa 1734 1591}%
\special{fp}%
%
\special{pn 8}%
\special{pa 2822 1596}%
\special{pa 2840 1591}%
\special{fp}%
\special{sh 1}%
\special{pa 2840 1591}%
\special{pa 2770 1590}%
\special{pa 2789 1605}%
\special{pa 2781 1628}%
\special{pa 2840 1591}%
\special{fp}%
\put(23.0000,-18.0000){\makebox(0,0){\LARGE $L$}}%
%
\special{pn 8}%
\special{pa 2900 1250}%
\special{pa 3100 1250}%
\special{fp}%
\special{sh 1}%
\special{pa 3100 1250}%
\special{pa 3033 1230}%
\special{pa 3047 1250}%
\special{pa 3033 1270}%
\special{pa 3100 1250}%
\special{fp}%
\special{pa 2500 1250}%
\special{pa 2300 1250}%
\special{fp}%
\special{sh 1}%
\special{pa 2300 1250}%
\special{pa 2367 1270}%
\special{pa 2353 1250}%
\special{pa 2367 1230}%
\special{pa 2300 1250}%
\special{fp}%
\put(27.0000,-12.5000){\makebox(0,0){\LARGE $R$}}%
%
\special{pn 15}%
\special{pa 1700 1380}%
\special{pa 1725 1540}%
\special{pa 1731 1572}%
\special{pa 1736 1604}%
\special{pa 1742 1636}%
\special{pa 1747 1667}%
\special{pa 1759 1731}%
\special{pa 1766 1762}%
\special{pa 1772 1794}%
\special{pa 1779 1825}%
\special{pa 1787 1856}%
\special{pa 1794 1887}%
\special{pa 1810 1949}%
\special{pa 1819 1979}%
\special{pa 1828 2010}%
\special{pa 1837 2040}%
\special{pa 1847 2070}%
\special{pa 1858 2100}%
\special{pa 1868 2130}%
\special{pa 1880 2160}%
\special{pa 1904 2218}%
\special{pa 1930 2276}%
\special{pa 1944 2304}%
\special{pa 1958 2333}%
\special{pa 2003 2417}%
\special{pa 2019 2445}%
\special{pa 2035 2472}%
\special{pa 2051 2500}%
\special{pa 2068 2527}%
\special{pa 2085 2555}%
\special{pa 2119 2609}%
\special{pa 2137 2636}%
\special{pa 2154 2663}%
\special{pa 2226 2771}%
\special{pa 2245 2798}%
\special{pa 2263 2825}%
\special{pa 2280 2850}%
\special{fp}%
%
\special{pn 15}%
\special{pa 2870 1380}%
\special{pa 2845 1540}%
\special{pa 2839 1572}%
\special{pa 2834 1604}%
\special{pa 2828 1636}%
\special{pa 2823 1667}%
\special{pa 2811 1731}%
\special{pa 2804 1762}%
\special{pa 2798 1794}%
\special{pa 2791 1825}%
\special{pa 2783 1856}%
\special{pa 2776 1887}%
\special{pa 2760 1949}%
\special{pa 2751 1979}%
\special{pa 2742 2010}%
\special{pa 2733 2040}%
\special{pa 2723 2070}%
\special{pa 2712 2100}%
\special{pa 2702 2130}%
\special{pa 2690 2160}%
\special{pa 2666 2218}%
\special{pa 2640 2276}%
\special{pa 2626 2304}%
\special{pa 2612 2333}%
\special{pa 2567 2417}%
\special{pa 2551 2445}%
\special{pa 2535 2472}%
\special{pa 2519 2500}%
\special{pa 2502 2527}%
\special{pa 2485 2555}%
\special{pa 2451 2609}%
\special{pa 2433 2636}%
\special{pa 2416 2663}%
\special{pa 2344 2771}%
\special{pa 2325 2798}%
\special{pa 2307 2825}%
\special{pa 2290 2850}%
\special{fp}%
\end{picture}}%
} 
\caption{The V-embedding in the confining background}
\label{VE}
\end{figure}
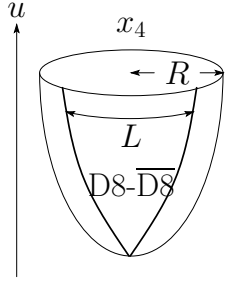

The asymptotic $\text{D8-}\overline{\text{D8}}$ distance $L$ can be evaluate by using (\ref{318}). The dependence of $L$ on $p_{2}$ evaluated numerically from (\ref{318}) and (\ref{319}) is shown by Fig.\ref{LCC}. 
At no electric field the only possible embedding in the confined phase is the U-embedding and $L$ becomes a decreasing monotonic function of $p_{2}$. However, the behavior of $L$ on $p_{2}$ can be modified under some external electric field. For $e^{2} > u_{K}^{3}$ the asymptotic behavior of $L$ becomes the same as in the deconfined phase. There is a threshold $e_{thr} (> u_{K}^{3/2})$ that modifies the behavior of $L$ on $p_{2}$ and the U-embedding exists for $e<e_{thr}$. For $e<e_{thr}$ the corresponding (on-shell) action is given by
\begin{align}
\label{321}
S^{\rm D8}_{\rm U} &={\cal N}\int d^{4}x du \; 
\dfrac{u^{5/2}}{\sqrt{f_{T}(u)}} 
\sqrt{\left(1 - \dfrac{e^{2}}{u^{3}} \right)
\Bigl(1- 9\mu^{2}\theta^{3}uh(u)^{2} \Bigr) } 
\nonumber \\
& \qquad \times \left[ 1- \dfrac{u_{0}^{8}f_{K}(u_{0})
\left(1 - \dfrac{e^{2}}{u_{0}^{3}} \right)
\Bigl(1 - 9\mu^{2}\theta^{3}uh(u_{0})^{2} \Bigr)}
{u^{8}f_{K}(u)\left(1 - \dfrac{e^{2}}{u^{3}} \right)
\Bigl(1 - 9\mu^{2}\theta^{3}uh(u)^{2} \Bigr)} 
\right]^{-1/2} \;.
\end{align}
The modification of the behavior of $L$ suggests the existence of another kind 
of $\text{D8-}\overline{\text{D8}}$ embedding in the confining background. The $\text{D8}$-brane and $\overline{\text{D8}}$-brane are adjusted in parallel and are connected at $u=u_{K}$ in this embedding. We refer to this embedding as ``V-shaped embedding" \cite{BLL}. (See Fig. \ref{VE})
In the V-embedding $\tau$ satisfies $\tau'(u)=0$ except at $u=u_{K}$ and its action is given by
\begin{align}
\label{322}
S^{\rm D8}_{\rm V} &={\cal N}\int d^{4}x du \; 
\dfrac{u^{5/2}}{\sqrt{f_{K}(u)}} 
\left\{1- \dfrac{3\mu\theta^{3/2}h(u) \widetilde{j}_{eT}}{u^{2}} \right\}
\sqrt{ \dfrac{1 - \dfrac{e^{2}}{u^{3}}}
{1 - \dfrac{\widetilde{j}_{eT}^{2}}{u^{5}}} } \;.
\end{align}
The reality condition for this action $S^{\rm D8}_{\rm V}$ in $e^{2} > u_{K}^{3}$ implies the existence of the nonvanishing current in the following form
\begin{align}
\label{323}
j_{eK}=e^{5/3}-\dfrac{3\mu\theta^{3/2}e^{2}}{1+\theta^{3}e^{2}} \;.
\end{align}
The V-embedding is therefore a conductor with conductivity
\begin{align}
\label{324}
\sigma_{K} =\dfrac{(2\pi\alpha')^{2}{\cal N}}{V_{4}}\dfrac{j_{eK}}{e} 
=\dfrac{N_{f}N_{c}\lambda}{48\pi^{3}}
\left[e^{2/3} - \dfrac{3\mu\theta^{3/2}e}{1+\theta^{3}e^{2}} \right]\;.
\end{align}
The conductivity also depends on the noncommutativity parameter as in the deconfinement phase and becomes the ordinary one in the limit of $\theta \to 0$ \cite{BLL}. 

\begin{figure}[H]
\begin{minipage}{0.5\hsize}
\centering
\includegraphics[width=60mm]{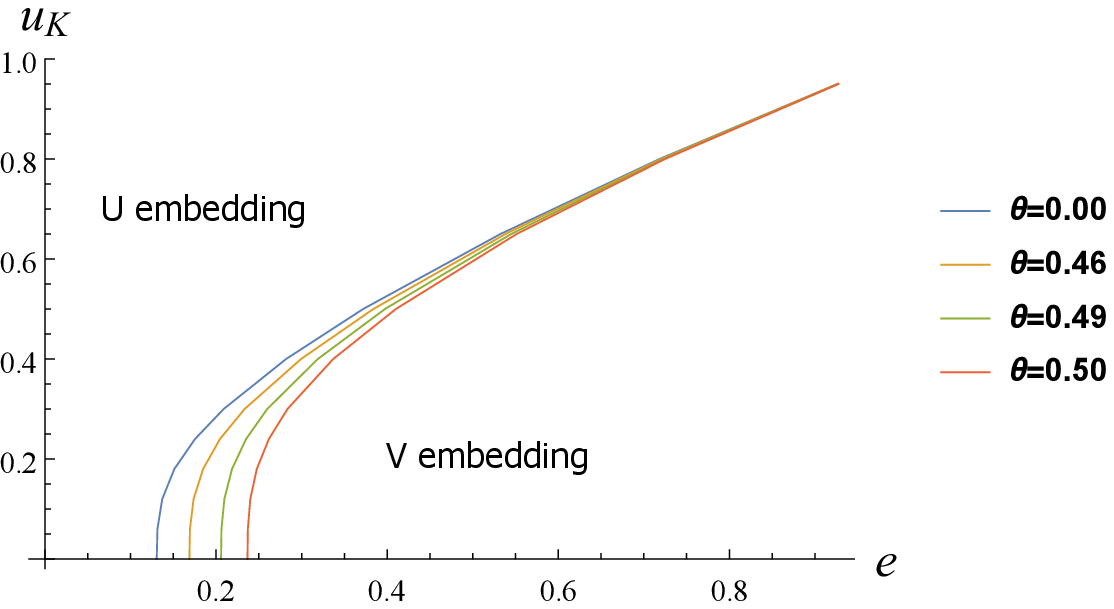} \\
$0.00 \leq \theta \leq 0.50$
\end{minipage}
\begin{minipage}{0.5\hsize}
\centering
\includegraphics[width=60mm]{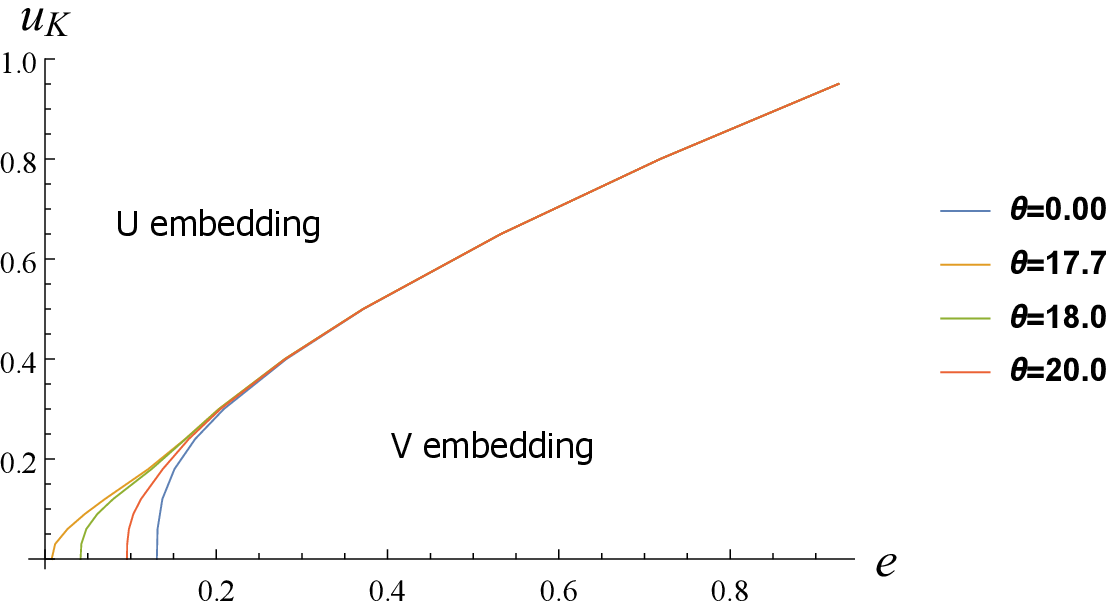} \\
$17.7 \leq \theta$
\end{minipage}
\caption{Phase diagram at finite parameter $u_{K}$ and electric field in the confining background. ($L=1$, $\mu=1$)}
\label{EKC}
\end{figure}

In the deconfinement phase, the current is produced due to the movement of 
quarks and anti-quarks, namely fundamental strings. In the confinement phase, the only charged objects are baryons. The current in the confinement phase can be regarded due to the movement of baryons and anti-baryons, namely $\text{D4}$-branes (and $\overline{\text{D4}}$-branes) wrapped on the $S^{4}$. It is thought that this stability of the cups singularity is provided by the balance of the forces caused by the $\text{D8}$-brane and the $\text{D4}$-branes pulling against each other. In accordance with this interpretation, we can evaluate the phase diagram in the $(u_{K},\;e)$ plane in the same way as  the deconfining phase. The phase diagram in the $(u_{K},\;e)$ plane with fixed the $\text{D8-}\overline{\text{D8}}$-brane distance and $\theta$ in the confining phase is shown in fig. \ref{EKC}. 

The global behavior of the phase diagrams has also not significantly changed  even at finite noncommutativity parameter $\theta$ in this situation, that is, the transition value of $u_{K}(=u_{Kc})$ increases as the transition electric field strength $e_{c}$ increases. Whereas $e_{c}$ at $T_{c}=0$ with finite $\theta$ is bigger than that with $\theta=0$ in the range of $0 \leq \theta \leq 0.50$, $e_{c}$ at $T_{c}=0$ with finite $\theta$ is smaller than that with $\theta=0$ in the range of $17.7 \leq \theta$. There is a tendency that $e_{c}$ is modified by $\theta$ as $u_{Kc}$ becomes smaller. We note that, even at finite noncommutativity parameter, $e_{c}$ in the limit $u_{Kc} \to 0$ is the same as that in the deconfinement phase in the limit $T \to 0$ \cite{BLL}. As $\theta$ approaches infinity, $e_{c}$ at $T_{c}=0$ turn back to them at zero noncommutativity parameter. The reality condition is not satisfied in the range of $0.50 < \theta < 17.7$ as in the deconfinement phase.

%
%

\section{Magnetic Field} 
\setcounter{equation}{0}
\addtocounter{enumi}{1}

We next investigate the response of the noncommutative deformation of the holographic QCD model at nonzero temperature to an external magnetic field $B$. We make an ansatz 
\begin{align}
\label{401}
2\pi \alpha'A_{0}=\mu \;, \quad 
2\pi \alpha'A_{1}(x_{2},\;u)=-bx_{2}+a_{1}(u)
\end{align}
where $\mu$ and $b$ are constants. As was seen in the previous section, the noncommutativity also have the effect of varying the transition magnetic field strength $b_{c}$.

\subsection{Deconfinement phase} 

Consider again the deconfining background, which dominates at high temperature $T >\dfrac{1}{2\pi R}$. The total action is given by
\begin{align}
\label{402}
S^{\rm D8} &=S^{\rm D8}_{\rm DBI} + S^{\rm D8}_{\rm WZ} \nonumber \\
&={\cal N}\int d^{4}x\,du \; \biggl[ u^{4} \sqrt{
\left(f_{T}(u)\tau'(u)^{2} + \dfrac{1}{u^{3}} \right)
\left(1 + \dfrac{b^{2}}{u^{3}} \right)
+ \dfrac{f_{T}(u)(a'_{1}(u))^{2}}{u^{3}} } \Biggr. \nonumber \\
& \hspace{30mm} \biggl. - 3\mu \theta^{3/2}u^{3}h(u)a_{1}'(u) \biggr]  
\nonumber \\
& ={\cal N}\int d^{4}x du \; u^{4} 
\left\{1-\dfrac{3\mu \theta^{3/2}h(u)\widetilde{j}_{bT}}{u^{2}f_{T}(u)} \right\}
\sqrt{\dfrac{\left(f_{T}(u)\tau'(u)^{2} + \dfrac{1}{u^{3}} \right)
\left(1 + \dfrac{b^{2}}{u^{3}} \right)}
{\left(1 - \dfrac{\widetilde{j}_{bT}^{2}}{u^{2}f_{T}(u)} \right)}} \;,
\end{align}
where $\widetilde{j}_{bT}=j_{bT}+3\mu\theta^{3/2}u^{3}h(u)$ with 
\begin{align}
\label{403}
j_{bT}=\dfrac{uf_{T}(u)a'_{1}(u)}{\sqrt{
\left(f_{T}(u)\tau'(u)^{2} + \dfrac{1}{u^{3}} \right)
\left(1 + \dfrac{b^{2}}{u^{3}} \right)
+ \dfrac{f_{T}(u)(a'_{1}(u))^{2}}{u^{3}} }}
- 3\mu \theta^{3/2}u^{3}h(u) \;.
\end{align}
The solution of the equation of motion and the constant of motion for $\tau(u)$ in the U-embedding (with the vanishing $j_{bT}$) are given respectively by
\begin{align}
\label{404}
\tau'(u) &=\dfrac{1}{u^{3/2}\sqrt{f_{T}(u)}} 
\left[\dfrac{u^{8}\left(1+\dfrac{b^{2}}{u^{3}}\right)
\Bigl(f_{T}(u)-9\mu^{2}\theta^{3}uh(u)^{2}\Bigr)}
{u_{0}^{8}\left(1+\dfrac{b^{2}}{u_{0}^{3}}\right)
\Bigl(f_{T}(u_{0})-9\mu^{2}\theta^{3}u_{0}h(u_{0})^{2}\Bigr)}-1 \right]^{-1/2} \;,
\end{align}
and 
\begin{align}
\label{405}
p_{3}=u_{0}^{4}\sqrt{f_{T}(u_{0})\left(1+\dfrac{b^{2}}{u_{0}^{3}}\right)
\Bigl(f_{T}(u_{0})-9\mu^{2}\theta^{3}u_{0}h(u_{0})^{2} \Bigr)}\;.
\end{align}
Although the reality conditions of the constant $p_{3}$ in (\ref{405}) has no restriction for the parameter $b$, it has same restriction as (\ref{310}) for the parameter $\theta$. 

In the U-embedding, the corresponding (on-shell) action without $j_{bT}$ is given by 
\begin{align}
\label{406}
S^{\rm D8}_{\rm U} &={\cal N}\int d^{4}x du \; 
\dfrac{u^{5/2}}{\sqrt{f_{T}(u)}} 
\sqrt{\left(1 + \dfrac{b^{2}}{u^{3}} \right)
\Bigl(f_{T}(u)- 9\mu^{2}\theta^{3}uh(u)^{2} \Bigr) } 
\nonumber \\
& \qquad \times \left[ 1- \dfrac{u_{0}^{8}
\left(1 + \dfrac{b^{2}}{u_{0}^{3}f_{T}(u_{0})} \right)
\Bigl(f_{T}(u_{0}) - 9\mu^{2}\theta^{3}uh(u_{0})^{2} \Bigr)}
{u^{8}\left(1 + \dfrac{b^{2}}{u^{3}f_{T}(u)} \right)
\Bigl(f_{T}(u) - 9\mu^{2}\theta^{3}uh(u)^{2} \Bigr)} 
\right]^{-1/2} \;.
\end{align}
In the parallel embedding with $\tau'(u)=0$, the action becomes
\begin{align}
\label{407}
S^{\rm D8}_{\rm ||} &={\cal N}\int d^{4}x du \; 
\dfrac{u^{5/2}}{\sqrt{f_{T}(u)}} 
\left\{f_{T}(u)- \dfrac{3\mu\theta^{3/2}h(u) \widetilde{j}_{bT}}{u^{2}} \right\}
\sqrt{ \dfrac{1 + \dfrac{b^{2}}{u^{3}}}
{f_{T}(u) - \dfrac{\widetilde{j}_{bT}^{2}}{u^{5}}} } \;.
\end{align}

\begin{figure}[H]
\begin{minipage}{0.5\hsize}
\centering
\includegraphics[width=60mm]{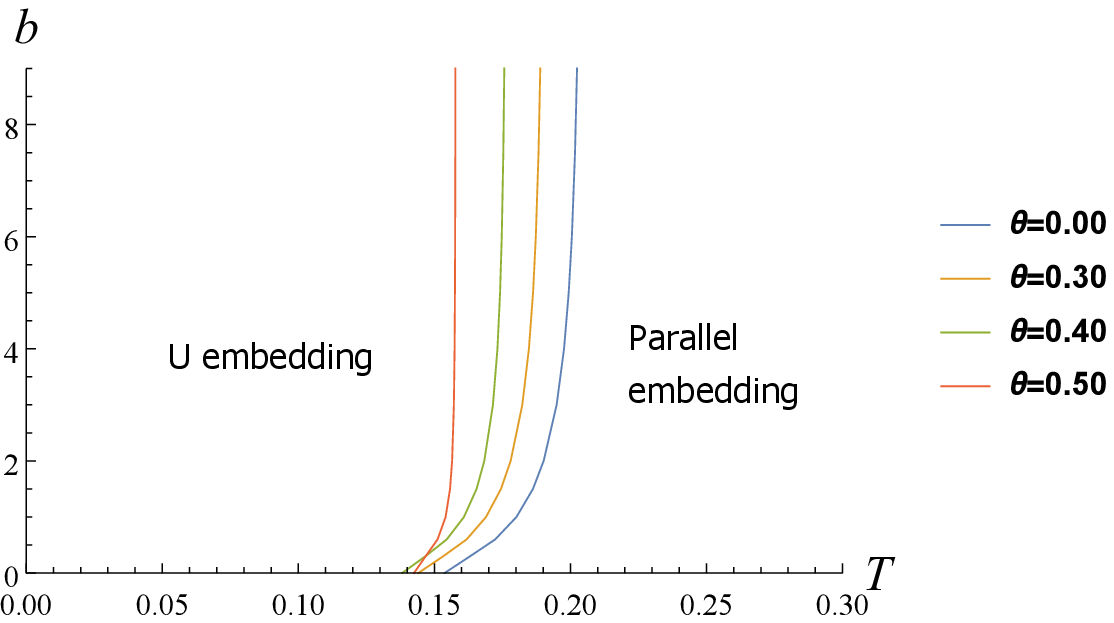} \\
$0.00 \leq \theta \leq 0.50$
\end{minipage}
\begin{minipage}{0.5\hsize}
\centering
\includegraphics[width=60mm]{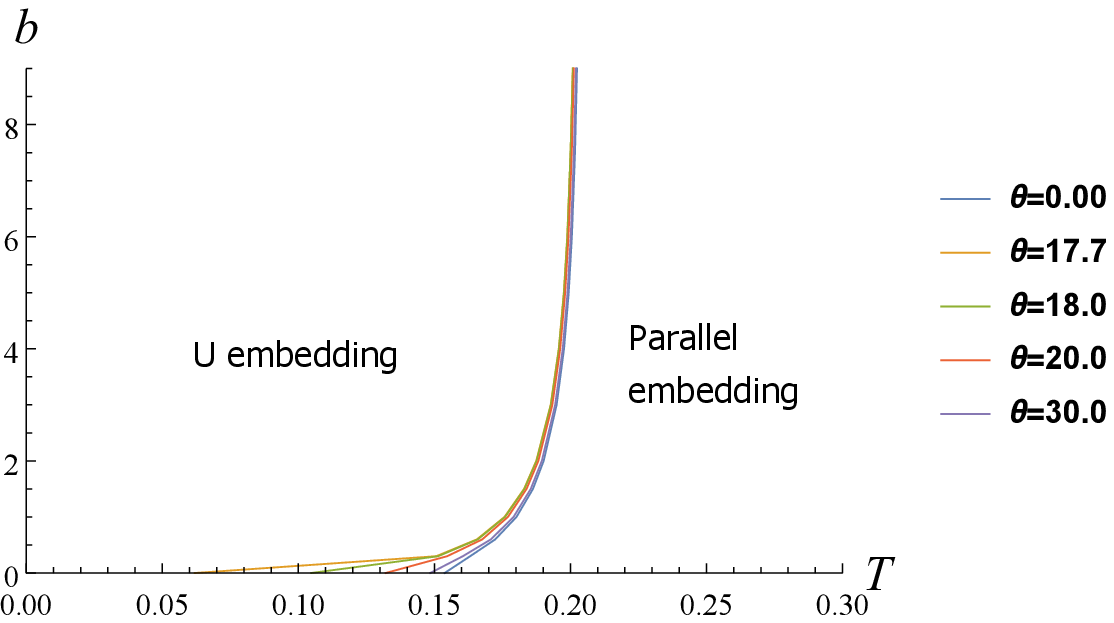} \\
$17.7 \leq \theta$
\end{minipage}
\caption{Phase diagram at finite temperature and magnetic field in the deconfining background. ($L=1$, $\mu=1$)}
\label{BTDC}
\end{figure}

We can determine the transition temperature $T_{c}$ and the transition magnetic  field strength $b_{c}$ by the Maxwell equal area law and construct the phase diagram in the $(b,\;T)$ plane with fixed $L$ and $\theta$. The phase diagram at nonzero temperature, background magnetic field and noncommutativity parameter in  the deconfining phase is shown in Fig.\ref{BTDC}. \\

The global behavior of the phase diagram has also not significantly changed even at  finite noncommutativity parameter $\theta$, that is, the transition temperature $T_{c}$ increases as the transition magnetic field strength $b_{c}$ increases even at finite noncommutativity parameter $\theta$. Although $T_{c}$ hardly changes at $b_{c}=0$, it significantly changes in the large-$b_{c}$ regime in the range of $0 \leq \theta \leq 0.50$. In contrast, although $T_{c}$ significantly changes at $b_{c}=0$, hardly changes in the large-$b_{c}$ regime in the range of $17.7 \leq \theta$. There's a tendency that $T_{c}$ decreases as $\theta$ increases in the range of $0 \leq \theta \leq 0.50$ and $T_{c}$ increases as $\theta$ increases in the range of $17.7 \leq \theta$. However, $T_{c}$ at $b_{c}=0$ with $\theta=0.40$ becomes smaller than that at $b_{c}=0$ with $\theta=0.50$. As $\theta$ approaches infinity, both $T_{c}$ and $b_{c}$ turn back to them at zero noncommutativity parameter. The reality condition is not satisfied in the range of $0.50 < \theta < 17.7$.

\subsection{Confinement phase} 

We next consider the confining background. In the U-embedding, the solution of the equation of motion for $\tau(u)$ and the constant of motion for $\tau(u)$ is given respectively by
\begin{align}
\label{408}
\tau'(u) &=\dfrac{1}{u^{3/2}f_{K}(u)} 
\left[\dfrac{u^{8}f_{K}(u)\left(1+\dfrac{b^{2}}{u^{3}}\right)
\Bigl(1-9\mu^{2}\theta^{3}uh(u)^{2}\Bigr)}
{u_{0}^{8}f_{K}(u_{0})\left(1+\dfrac{b^{2}}{u_{0}^{3}}\right)
\Bigl(1-9\mu^{2}\theta^{3}u_{0}h(u_{0})^{2}\Bigr)}-1 \right]^{-1/2} \;,
\end{align}
and 
\begin{align}
\label{409}
p_{4}=u_{0}^{4}\sqrt{f_{K}(u_{0})\left(1+\dfrac{b^{2}}{u_{0}^{3}}\right)
\Bigl(1-9\mu^{2}\theta^{3}u_{0}h(u_{0})^{2} \Bigr)}\;.
\end{align}
The solution $\tau'(u)$ and the constant of motion $p_{4}$ are same as in (\ref{318}) and in (\ref{319}) with substitution of $-e^{2}$ for $b^{2}$, respectively. Due to the difference in sign, the asymptotic $\text{D8-}\overline{\text{D8}}$ distance $L$ is a decreasing monotonic function of $p_{4}$ for all values of $b$. It can be concluded that the only possible embedding in the confined phase is the U-embedding. The on-shell action in the U-embedding is given by
\begin{align}
\label{410}
S^{\rm D8}_{\rm U} &={\cal N}\int d^{4}x du \; 
\dfrac{u^{5/2}}{\sqrt{f_{T}(u)}} 
\sqrt{\left(1 + \dfrac{b^{2}}{u^{3}} \right)
\Bigl(1- 9\mu^{2}\theta^{3}uh(u)^{2} \Bigr) } 
\nonumber \\
& \qquad \times \left[ 1- \dfrac{u_{0}^{8}f_{K}(u_{0})
\left(1 + \dfrac{b^{2}}{u_{0}^{3}} \right)
\Bigl(1 - 9\mu^{2}\theta^{3}uh(u_{0})^{2} \Bigr)}
{u^{8}f_{K}(u)\left(1 + \dfrac{b^{2}}{u^{3}} \right)
\Bigl(1 - 9\mu^{2}\theta^{3}uh(u)^{2} \Bigr)} 
\right]^{-1/2} \;.
\end{align}
%

%
%
\section{Conclusions and Discussions}
\setcounter{section}{4}
\setcounter{equation}{0}
\addtocounter{enumi}{1}

In this paper, we have constructed a noncommutative deformation of the 
holographic QCD (Sakai--Sugimoto) model at finite temperature in accordance with a prescription of \cite{HI, MR, AOSJ, APR} and have examined the response to external electric and magnetic fields regarding baryon number currents by this model. The noncommutative deformation of the gauge theory does not change the phase structure with respect to the baryon number current. There is also the conductor phase in addition to the insulator phase even in the noncommutative deformation of the confinement background at finite electric field \cite{BLL}. However, the transition temperature $T_{c}$, the transition electric field $e_{c}$ and magnetic field $b_{c}$ in the conductor-insulator phase transition depend on the noncommutativity parameter $\theta$. Namely, the noncommutativity of space coordinates has an influence on the shape of the phase diagram for the conductor-insulator phase transition. It is known that the noncommutativity of space coordinates  also has an influence on the shape of the phase diagram for the chiral symmetry breaking-chiral symmetry restoration within the framework of the noncommutative deformation of the holographic QCD model at finite temperature \cite{TN_YO_KS}. It can be regarded as an example that the noncommutativity of space coordinates reflects physical quantities \cite{TN_YO_KS2, NST}. 

The phase diagrams have shown that the transition temperature $T_{c}$, the transition electric field $e_{c}$ and magnetic field $b_{c}$ shift to the commutative ones in the zero noncommutativity parameter limit. On the contrary, the phase diagrams have shown that $T_{c}, e_{c}$ and $b_{c}$ also shift to the commutative ones in the infinite noncommutativity parameter limit. It can be easily seen that the nonvanishing currents $j_{eT},\;j_{eK}$ and the conductivities $\sigma_{T},\;\sigma_{K}$ reduce to the commutative ones in both the zero and infinite noncommutativity parameter limit. These properties are suggestive to a kind of the Morita duality between irreducible modules over the noncommutative torus \cite{DN, APR}. On the other hand, the allowed range of the noncommutativity parameter can be restricted by the reality condition of the constants of motion. It might be remarkable that the restriction on the noncommutativity parameter by the physical conditions. 

In the holographic QCD model, a chemical potential for baryon number corresponds
to a nonzero asymptotic value of the electrostatic potential on the D8-branes 
\cite{KSZ, HT, PS2, BLL2, Y, RSRW, KSZ2}. In our model, a constant baryon chemical potential has been naively introduced. The dependence of $T_{c}, e_{c}$ and $b_{c}$ on the baryon chemical potential should be considered in detailed procedures.

An alternative gravity dual of the confinement-deconfinement phase transition in the Sakai--Sugimoto model has been proposed in \cite{MM1,MM2,IMM}. Ref. \cite{MM1} have argued that the gravity dual of the deconfinement transition is a Gregory-Laflamme transition into the T-dual type IIB supergravity, where the black D4-brane geometry is replaced by an localized D3 brane geometry. It would be interesting to study the properties of the baryon number current in this model. 

The UV/IR mixing is well known as distinctive features of noncommutative field 
theories. The phenomenon of the UV/IR mixing appears to be the qualitative difference between ordinary and noncommutative field theory. The difference in the properties of the baryon number current between ordinary and noncommutative QCD might be related to the UV/IR mixing. We hope to discuss this subject in the future.

\section*{Acknowledgments}

We are grateful to C. N\'u\~nez for useful discussions and comments. One of us (T.N.) would like to thank members of the Physics Department at College of Engineering, Nihon University for their encouragements. This work was supported in part by the overseas research fund of Nihon university. 

\clearpage 

%
%

\end{document}